\documentclass[final,12pt]{elsarticle}
\usepackage[numbers]{natbib}
\usepackage{amssymb}
\usepackage{amsmath}
\usepackage{mathtools}
\usepackage{enumitem}
\usepackage[T1]{fontenc}
\usepackage[utf8]{inputenc}
\usepackage{lipsum}
\usepackage{lineno}
\usepackage{hyperref}
\usepackage{multirow}
\usepackage{graphicx}
\usepackage{subcaption}
\usepackage[table,xcdraw]{xcolor}
\usepackage{booktabs}
\usepackage{float}
\usepackage{cleveref}
\usepackage{algorithm,algpseudocode}

\newcommand{\bfI}{{\bf I}}

\newcommand{\bfK}{{\bf K}}

\newcommand{\bfQ}{{\bf Q}}

\newcommand{\bfV}{{\bf V}}

\newcommand{\bfs}{{\bf s}}
\newcommand{\bfg}{{\bf g}}
\newcommand{\bfx}{{\bf x}}

\newcommand{\bfp}{{\bf p}}

\DeclareMathOperator*{\argmin}{argmin}
\DeclareMathOperator*{\argmax}{argmax}

\journal{Computerized Medical Imaging and Graphics}

\begin{document}

\begin{frontmatter}

\title{IgCONDA-PET: Weakly-Supervised PET Anomaly Detection using Implicitly-Guided Attention-Conditional Counterfactual Diffusion Modeling -- a Multi-Center, Multi-Cancer, and Multi-Tracer Study}

\author[afl1,afl2]{Shadab Ahamed}\corref{cor1} 
\author[afl1,afl2,afl3]{Arman Rahmim}
\cortext[cor1]{Corresponding author: Shadab Ahamed (email: shadab.ahamed@hotmail.com)}

\affiliation[afl1]{organization={Department of Physics \& Astronomy, University of British Columbia},
            city={Vancouver},
            state={BC},
            country={Canada}}
\affiliation[afl2]{organization={Department of Integrative Oncology, BC Cancer Research Institute},
            city={Vancouver},
            state={BC},
            country={Canada}}
\affiliation[afl3]{organization={Department of Radiology, University of British Columbia},
            city={Vancouver},
            state={BC},
            country={Canada}}
\begin{abstract}
Minimizing the need for pixel-level annotated data to train PET lesion detection and segmentation networks is highly desired and can be  transformative, given time and cost constraints associated with expert annotations. Current unsupervised or weakly-supervised anomaly detection methods rely on autoencoder or generative adversarial networks (GANs) trained only on healthy data. While these approaches reduce annotation dependency, GAN-based methods are notably more challenging to train than non-GAN alternatives (such as autoencoders) due to issues such as the simultaneous optimization of two competing networks, mode collapse, and training instability. In this paper, we present the weakly-supervised \textbf{I}mplicitly \textbf{g}uided \textbf{CO}u\textbf{N}terfactual diffusion model for \textbf{D}etecting \textbf{A}nomalies in \textbf{PET} images (IgCONDA-PET). The solution is developed and validated using PET scans from six retrospective cohorts consisting of a total of 2652 cases (multi-cancer, multi-tracer) containing both local and public datasets (spanning multiple centers). The training is conditioned on image class labels (healthy vs.~unhealthy) via attention modules, and we employ implicit diffusion guidance. We perform counterfactual generation which facilitates ``unhealthy-to-healthy'' domain translation by generating a synthetic, healthy version of an unhealthy input image, enabling the detection of anomalies through the calculated differences. The performance of our method was compared against several other deep learning based weakly-supervised or unsupervised methods as well as traditional methods like 41\% SUV$_\text{max}$ thresholding. We also highlight the importance of incorporating attention modules in our network for the detection of small anomalies. The code is publicly available at: \url{https://github.com/ahxmeds/IgCONDA-PET.git}.
\end{abstract}




\end{frontmatter}

\section{Introduction}
\label{sec:introduction}
Detection of cancerous anomalies from positron emission tomography (PET) images is a critical step in the clinical workflow for oncology, aiding in treatment planning, radiotherapy, and surgical interventions \cite{scott2008pet,acuff2018practical,molina2009intra}. Oncological PET scans provide valuable metabolic information that helps in distinguishing malignant tissues from normal tissues, but the process of manual segmentation is prone to many challenges. Expert voxel-level annotation, while considered the gold standard, is not only time-consuming \cite{ahamed2022cascaded,yousefirizi2022convolutional} but also susceptible to intra- and inter-observer variability \cite{clinical_metrics_paper}, which can introduce inconsistencies and compromise the reliability of downstream analyses. This issue is exacerbated in large-scale studies and/or overburdened clinical settings where annotators must process numerous scans, increasing the potential for fatigue and error. As a result, Computer-Aided Detection (CADe) systems \cite{firmino2016computer} are emerging as valuable tools, enhancing the efficiency and accuracy of lesion detection while reducing reliance on manual annotation.

Recent advancements in deep learning and machine learning have paved the way for weakly-supervised approaches in medical anomaly detection \cite{what_is_healthy,ni2023weakly,hibi2023automated,diff_model_med_anomaly_detection}. Weakly supervised techniques are particularly promising for medical imaging applications, where the scarcity of detailed labeled data is a well-recognized challenge \cite{misera2024weakly}. These methods leverage image-level labels, which are significantly easier and quicker to obtain compared to dense pixel-level annotations, thereby addressing the time and resource constraints inherent in traditional segmentation workflows. Despite the impressive progress in fully-supervised PET lesion segmentation, its clinical translation is hampered by a fundamental data bottleneck: there are still very few publicly available oncology PET datasets that include reliable voxel-level ground-truth masks \cite{autopet_paper,hecktor_paper}. Even though physicians usually agree on which axial slices contain the disease, there can be noticeable variations in the exact placement of lesion boundaries by different physicians for the same image \cite{clinical_metrics_paper}.

\begin{figure}[H]
\centering
\includegraphics[width=0.7\textwidth]{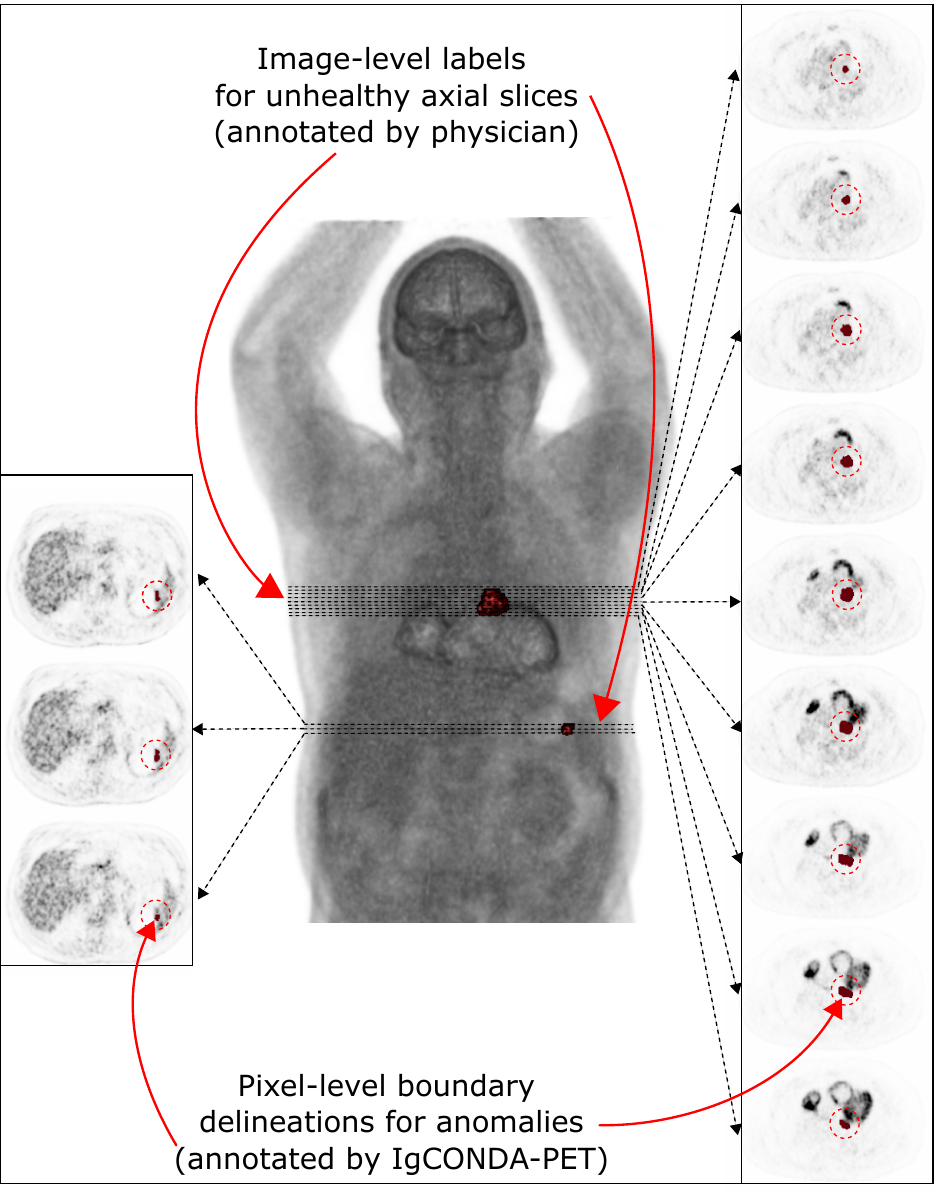}
\caption{\textbf{Weakly-supervised PET anomaly detection.} Rather than tracing labor-intensive and time-consuming voxel-level contours, the physician can simply flag the axial PET slices that are likely to contain pathologies. A weakly-supervised algorithm then converts these coarse slice-level cues into precise 3D lesion masks, yielding faster, more reproducible annotations and a richer supply of ground-truth labels for training and validation of deep neural networks. The lesions are shown in maroon inside red dashed circles on the axial slices.}
\label{fig:introduction_image}
\end{figure}

Weakly-supervised anomaly detection approaches trained only with coarse, slice-level (or study-level) labels therefore provide a practical alternative. As shown in \Cref{fig:introduction_image}, they exploit the consistent skill of lesion localization on axial slices by of experts, while delegating the tedious, fine-grained contouring task to an automated algorithm, yielding faster and more reproducible delineations \cite{what_is_healthy,ni2023weakly,djahnine2024weakly}. In turn, this lowers the annotation burden, enabling the rapid curation of much larger multi-center datasets, boosting statistical power and model generalizability across scanners, tracers, and cancer types. Because weak supervision can be harvested from routine clinical reports, it also eases privacy concerns around releasing detailed masks and supports continual learning from real-world data streams \cite{eyuboglu2021multi}. Finally, pixel-level predictions derived from weak labels can be plugged directly into CADe/CADx pipelines to flag subtle lesions, assist treatment-planning workflows, and standardize quantitative biomarkers \cite{clinical_metrics_paper,dzikunu2025comprehensive,dzikunu2025adaptive}. Collectively, these advantages make weakly-supervised PET anomaly detection an essential step toward scalable, trustworthy, and widely deployable oncologic imaging AI.


In this study, we exploit a weakly-supervised approach using a diffusion probabilistic model (DPM) for pixel-level anomaly detection in PET images. DPM, with their ability to capture complex data distributions, are uniquely suited for detecting subtle and small anomalies that may elude simpler models \cite{bhosale2024anomaly}. By combining weakly-supervised learning with the robust generative capabilities of DPMs, our approach aims to provide accurate and reliable pixel-level anomaly detection by just using the image-level labels as ground truth, mitigating the limitations of traditional methods while maintaining clinical relevance.

\begin{figure}
\centering
\includegraphics[width=1\textwidth]{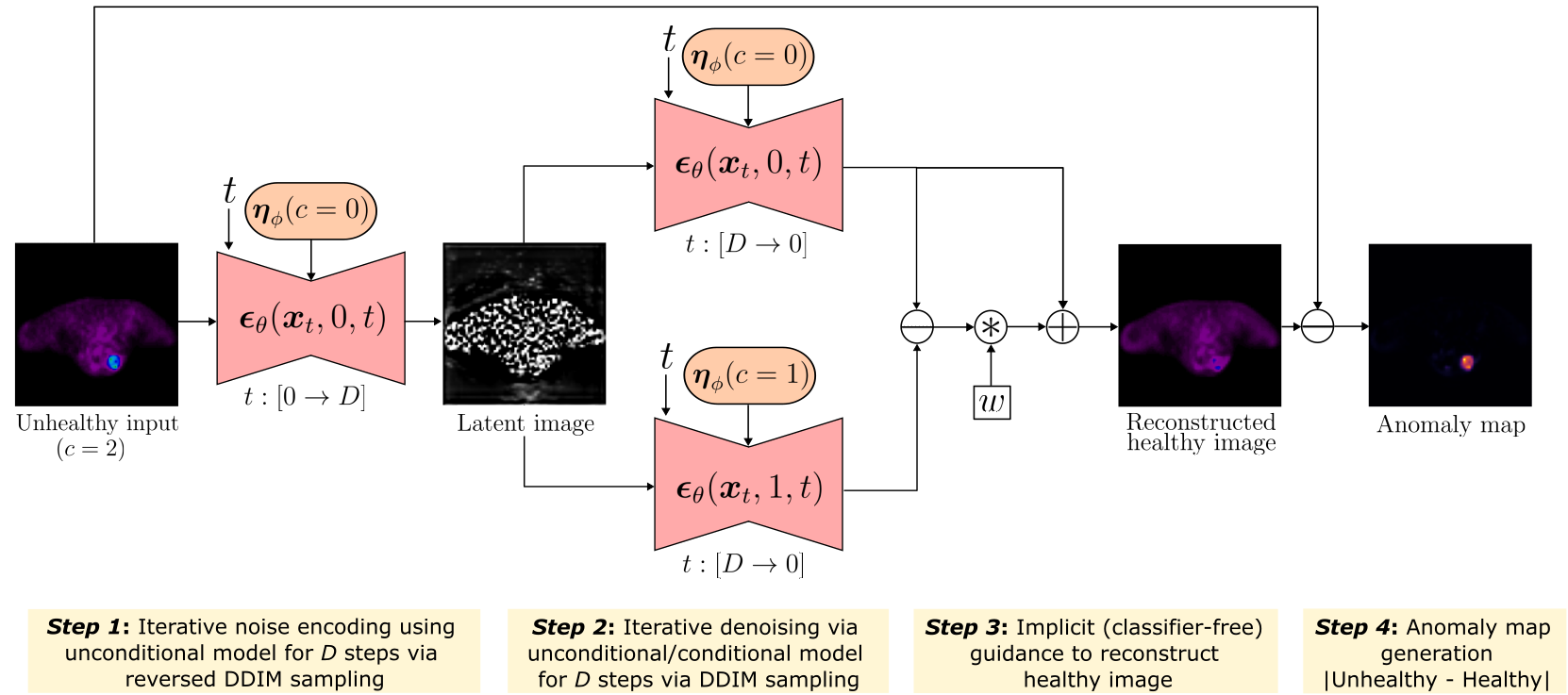}
\caption{\textbf{IgCONDA-PET:} Implicitly-guided counterfactual DPM methodology for domain translation between an unhealthy image and its healthy counterfactual for PET. The anomaly map is defined as the absolute difference between the unhealthy and corresponding reconstructed healthy image. Here, $\boldsymbol{\epsilon}_\theta$ and $\boldsymbol{\eta}_\phi$ represent the 3-level diffusion model UNet with attention mechanism and learnable class-embedding module parametrized by $\theta$ and $\phi$, respectively. $c=0, 1$, and $2$ denote unconditional, healthy and unhealthy class-conditioning labels, respectively.} 
\label{fig:method_scheme}
\end{figure}

\textbf{Related work.} Unsupervised deep learning-based anomaly detection in PET images has been explored in \cite{brainpet_17vae_comparison,Choi2019-ys,Hassanaly_2024}, although these were developed on brain PET datasets for anomalies related to dementia. Moreover, these methods were trained only on healthy cases under the assumption that since the model is trained to reconstruct only healthy data, it would fail on unhealthy cases in the regions of anomalies, thereby highlighting the unhealthy areas. Despite this simple idea, these models might not work well in practice because a lesion would deform regions around it and these deformations should not be captured by the anomaly detection algorithm. Moreover, as shown in \cite{baur2021autoencoders,what_is_healthy}, detecting anomalies without being shown examples of unhealthy data is non-trivial and such models often simply highlight regions of hyper-intensity in the image. Recently, diffusion models have been employed for medical anomaly detection \cite{diff_model_med_anomaly_detection,what_is_healthy}, but these have largely been validated only on brain MRI datasets alone. PET-based application of diffusion models have been explored in the context of image denoising \cite{pet_denoising_diffusion1,pet_denoising_diffusion2} and reconstruction \cite{pet_recon_diffusion1,pet_recon_diffusion2}, although their application to anomaly detection, especially in oncological PET use-cases have been limited \cite{dong2024head}. 

In this work, we propose a counterfactual DPM based on \cite{what_is_healthy}, trained on healthy and unhealthy axial PET slices with image-level labels. The class labels were preprocessed using an embedding module and were then fed into each level of the model augmented with attention mechanism \cite{class_cond_unet} (\Cref{subsec:attention_based_class_conditional_diffusion_model}). During inference, the synthesis process can be controlled via class labels and the anomalies were highlighted by conducting minimal intervention (known as counterfactual generation \cite{sanchez2022diffusion}) to perform an unhealthy to healthy domain translation. We then generate heatmaps by computing the difference between the unhealthy image and its reconstructed healthy counterfactual (\Cref{subsec:counterfactual_generation_and_anomaly_detection}). 

\textit{Contributions:} To the best of our knowledge, this is the first work on (i) counterfactual DPM for weakly-supervised PET anomaly detection using multi-institutional, multi-cancer and multi-tracer datasets. We (ii) train our models using implicit guidance (\Cref{subsec:implicit_guidance}), which eliminates the reliance on a downstream classifier for guidance (see, \Cref{subsec:implicit_guidance}) \cite{diff_model_med_anomaly_detection}; (iii) conduct extensive ablation studies with respect to the presence or absence of attention mechanism within the different levels of DPM network (\Cref{subsec:attention_based_class_conditional_diffusion_model}); (iv) perform experiments highlighting the sensitivity of the method to different inference hyperparameter choices; (v) show the superiority of our method against several other related state-of-the-art methods for weakly-supervised/unsupervised anomaly detection (\Cref{subsec:benchmark}) using slice-level metrics such as optimal Dice similarity coefficient (DSC) and 95\%tile Hausdorff distance (HD95), pixel-level metrics such as the area under the precision-recall curve (AUPRC), and lesion-level metrics such as the lesion detection sensitivity (\Cref{subsec:evaluation_metrics}). 

\section{Method}
\label{sec:method}
\subsection{Diffusion modeling}
\label{subsec:diffusion_model_a_primer}
Diffusion models are a class of generative models that rely on learning to reverse a diffusion process - typically a sequence of transformations that gradually add noise to data - to generate samples from noise. They consist of two main processes: a forward (noising) process and a reverse (denoising) process.

\textbf{Forward process (Noising):} The forward process in a diffusion model is a fixed Markovian process that iteratively adds Gaussian noise to the input (clean) image $\bfx_0 \sim p_\text{data}$ over a sequence of time steps $t \in \{1, 2, \ldots, T\}$ following a variance schedule $\beta_1, \beta_2, \ldots, \beta_T$ \cite{ddpm_paper}. For each time step $t$, noise is added to the image, resulting in a progressively noisier version of the original image. The transition of the image $\bfx_0$ at $t=0$ to $\bfx_t$ at time $t$ is governed by the distribution, 
\begin{equation}
\label{eq:forward_process_distribution}
    q(\bfx_t \mid \bfx_0) = \mathcal{N}(\bfx_t; \sqrt{\bar{\alpha_t}} \bfx_0, (1 - \bar{\alpha}_t) \bfI),
\end{equation}
where $\alpha_t = (1 - \beta_t)$ and $\bar{\alpha_t} = \prod_{i=0}^t \alpha_i = \prod_{i=0}^t (1 - \beta_i)$ denotes the cumulative effect of noise up to time $t$, representing how much signal from the original image remains after $t$ steps of noise addition. Here, the image $\bfx_t$ is given by,
\begin{equation}
\label{eq:forward_process_sampling}
    \bfx_t = \sqrt{\bar{\alpha}}_t \bfx_0 + \sqrt{1 - \bar{\alpha}_t} {\boldsymbol{\epsilon}},
\end{equation}
where the noise $\boldsymbol{\epsilon} \sim \mathcal{N}(0, \bfI)$ is sampled from a standard normal distribution. 

\textbf{Reverse process (Denoising):} The reverse process, which is learned during training, a network parametrized by $\theta$ learns to iteratively remove the added noise from the noisy image $\bfx_t$ recovering the original image $\bfx_0$. We denote the learned network for prediction of noise as $\boldsymbol{\epsilon}_\theta(\bfx_t, t)$. The reverse process learns a denoising distribution $p_\theta(\bfx_{t-1} \mid \bfx_t)$ that predicts the denoised image $\bfx_{t-1}$ parametrized by learnable $\theta$, modeled as,
\begin{equation}
\label{eq:reverse_process_distribution}
    p_\theta(\bfx_{t-1} \mid \bfx_t) = \mathcal{N}(\bfx_{t-1}; \boldsymbol{\mu}_\theta(\bfx_t, t), \beta_t \bfI),
\end{equation}
where 
\begin{equation}
\label{eq:reverse_process_mean}
    \boldsymbol{\mu}_\theta (\bfx_t, t) = \frac{1}{\sqrt{\alpha_t}}\bigg(\bfx_t - \frac{1 - \alpha_t}{\sqrt{1 - \bar{\alpha}_t}} \boldsymbol{\epsilon}_\theta(\bfx_t, t) \bigg),
\end{equation}  
is the mean predicted by the denoising network \cite{ddpm_paper}. During training of the denoising network, the Mean Squared Error (MSE) between the true added noise $\boldsymbol{\epsilon}$ and predicted noise $\boldsymbol{\epsilon}_\theta(\bfx_t, t)$ is minimized to find the optimal set of parameters $\theta^\star$. The training objective is given by, 
\begin{equation}
\label{eq:training_objective}
\theta^\star = \argmin_{\theta} \mathbb{E}_{\bfx_0, \boldsymbol{\epsilon}, t}\Big[\| \boldsymbol{\epsilon} - \boldsymbol{\epsilon}_\theta(\bfx_t, t) \|_2^2\Big]
\end{equation}
\textbf{Inference (Generation):} During inference or generation step, the trained model is iteratively applied on a noisy input to generate an image from the target distribution. As the DDPM sampling is stochastic \cite{ddpm_paper, chan2024tutorial} and requires $T$ denoising steps (where $T$ can be large),  Denoising Diffusion Implicit Models (DDIM) sampling \cite{ddim_paper} is often exploited for faster sampling. DDIM introduces a deterministic non-Markovian update rule which allows for reducing the number of steps during generation (due to the reparametrization trick, as explained in \cite{ddim_paper}) allowing for faster sampling while maintaining sample quality. For any pair of time steps $t$ and $t-k$, the DDIM update rule is given by, 
\begin{equation}
\label{eq:ddim_sampling}
    \bfx_{t-k} = \sqrt{\bar{\alpha}_{t-k}}\bigg(\frac{\bfx_t - \sqrt{1 - \bar{\alpha}_t} \boldsymbol{\epsilon}_\theta(\bfx_t, t) }{\sqrt{\bar{\alpha}_t}} \bigg) + \sqrt{1 - \bar{\alpha}_{t-k}} \boldsymbol{\epsilon}_\theta(\bfx_t, t).
\end{equation}

We used the DDIM sampler because it offers two crucial benefits that align directly with our counterfactual diffusion framework for PET lesion localization. First, DDIM provides an almost invertible, deterministic mapping between the fully noised latent $\bfx_T$ and the clean image $\bfx_0$, which lets us encode a PET slice into its unconditional latent state and then decode that exact latent under a guided noise schedule, guaranteeing pixel-wise correspondence between the original and counterfactual images - a property crucial for the final heatmap generation. Standard techniques such as DDPM sampling injects fresh Gaussian noise at every step, thereby producing a stochastic and therefore non-unique decoding path, which introduces speckle artifacts that confound lesion saliency. Second, DDIM can traverse the diffusion trajectory with a much coarser timestep grid (which is obtained by choosing a different under-sampling $t$ in $[0, T]$), yielding high-quality samples in much fewer number of steps as compared to DDPM. This  speedup cuts inference time from minutes to seconds, making the tool practical for real-time clinical integration while also reducing GPU cost during large-scale training and ablation.

\subsection{Attention-based class-conditional diffusion model}
\label{subsec:attention_based_class_conditional_diffusion_model}
In this work, we implemented diffusion modeling using a conditional denoising UNet $\boldsymbol{\epsilon}_\theta(\bfx_{t}, c, t)$, where the generation process could be controlled via the class labels $c$ of the images. Hence, for our use-case, we can replace $\boldsymbol{\epsilon}_\theta(\bfx_t, t)$ with $\boldsymbol{\epsilon}_\theta(\bfx_t, c, t)$ in \Cref{eq:reverse_process_mean,eq:training_objective,eq:ddim_sampling}. The updated training objective for this network is given by, 
\begin{equation}
\label{eq:training_objective_with_class_condition}
    \theta^\star = \argmin_{\theta} \mathbb{E}_{\bfx_0, \boldsymbol{\epsilon}, t}\Big[\| \boldsymbol{\epsilon} - \boldsymbol{\epsilon}_\theta(\bfx_t, c, t) \|_2^2\Big]
\end{equation}
The class label $c$ was incorporated into the denoising UNet via attention mechanism that has shown to improve performance in \cite{ddpm_paper,classifier_guidance_paper,class_cond_unet,what_is_healthy}. We used an embedding layer $\boldsymbol{\eta}_{\phi}$(c) parametrized by trainable parameters $\phi$ of dictionary size $s$ and embedding dimension $d$ to project the class tokens into vector representation. These vector representations were fed into the UNet augmented with attention layers at each level. The attention modules were implemented as,
\begin{equation}
\label{eq:attention_mechanism}
    \text{Attention}(\bfQ, \bfK_c, \bfV_c) = \text{softmax}\bigg( \frac{\bfQ \bfK_c^\top}{\sqrt{d}}\bigg) \bfV_c,
\end{equation}
where $\bfQ$ is the query matrix, $\bfK_c = \text{concat}[\bfK, \boldsymbol{\eta}_\phi(c)]$ and $\bfV_c = \text{concat}[\bfV, \boldsymbol{\eta}_\phi(c)]$ are the augmented key and values matrices respectively, where $\bfQ, \bfK$ and $\bfV$ were derived from the previous convolutional layers \cite{ddpm_paper,classifier_guidance_paper}. 

Our UNet consisted of 3 resolution levels with 64 channels and one ResNet block \cite{he2016deep} or ResNet+Spatial-Transformer block \cite{jaderberg2015spatial} per level, as we explain later. Each level of UNet could incorporate attention mechanism with 16 channels per attention head. Hence, we denote our models using a 3-tuple $(k_1k_2k_3)$, where $k_i \in \{0, 1\}$, with 0 and 1 representing the absence and presence of attention, respectively, in the $i^{\text{th}}$ level. We ablated over three different model types, namely $(k_1k_2k_3) = (000), (001)$ and $(011)$, to gauge the benefit of adding attention progressively deeper in the hierarchy. To the best of our knowledge, this is the first work based on counterfactual DPM studying the effect of attention mechanism in different levels of UNet on PET anomaly detection performance. 

\begin{figure}[H]
\centering
\includegraphics[width=\textwidth]{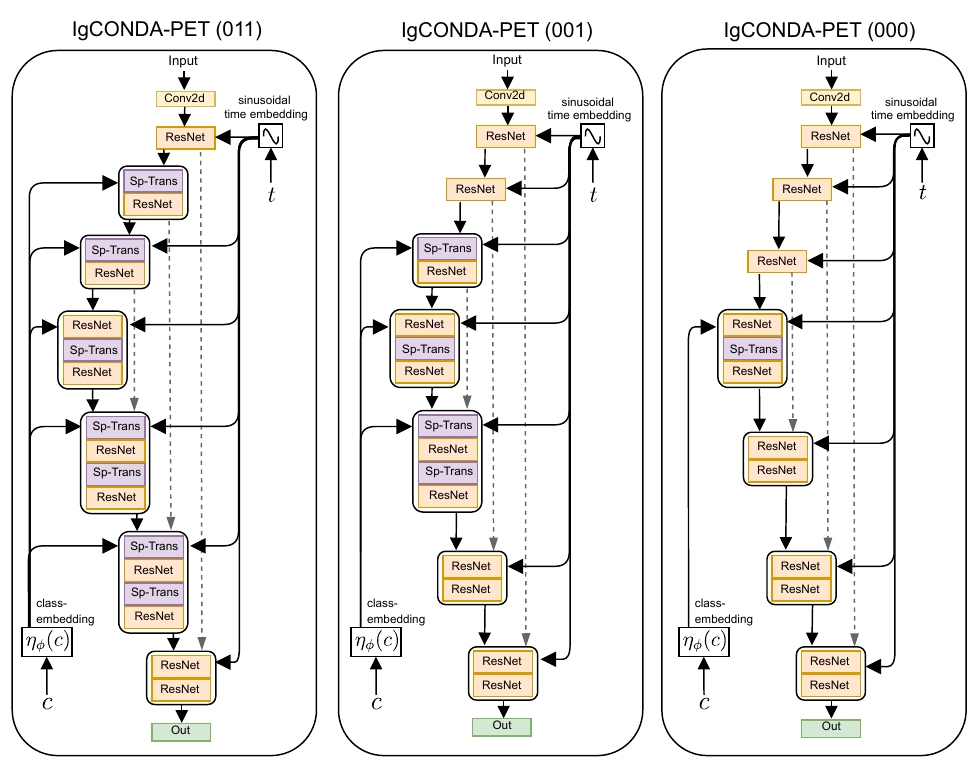}
\caption{\textbf{The three attention-variants of IgCONDA-PET -- (011) (left), (001) (middle) and (000) (right)}. The 3-level denoising diffusion UNet denoted by $\boldsymbol{\epsilon}_\theta$ takes a sinusoidal time-embedding of time $t$ and a class-embedding corresponding to class label $c$ parametrized by $\boldsymbol{\eta}_\phi$. The class-embedding is incorporated into the network via the Spatial-Transformers which implement the cross-attention mechanism to learn the association between the input image and the class label. In the 3-tuple $(k_1k_2k_3)$ representation, each occurrence of $k_i = 1$ denotes that the ResNet blocks in level $i$ has been replaced by a combination of ResNet+Spatial-Transformer (denoted as Sp-Trans in the figures) in both the downsampling and upsampling paths. The bottleneck layer in each network consists of a Spatial-Transformer sandwiched between two ResNet blocks.}
\label{fig:three_variants_of_igcondapet}
\end{figure}

We now discuss the network architecture with respect to the incorporation of attention in more detail. For any level $i$, setting $k_i = 1$ upgrades all the occurrences of pure ResNet ($k_i = 0$) to a ResNet+Spatial-Transformer block. The Spatial-Transformer block is a lightweight Vision Transformer inserted immediately after a ResNet block from the previous level. It (i) projects the feature map with a $1\times 1$ convolution; (ii) flattens the feature map grid into a sequence of tokens; (iii) applies two attention operation - self-attention among the tokens and cross-attention to the class vector $c$ - followed by a GEGLU \cite{shazeer2020glu} feed-forward layer; and (iv) reshapes the tokens back to feature map form and adds them element-wise to the original ResNet features that entered the Spatial-Transformer, closing the residual path. Because the same module is used on the encoder, at the bottleneck and on the decoders paths, global class information can influence feature maps at every stage of the network. Schematics for the different attention-variants of IgCONDA-PET are presented in \Cref{fig:three_variants_of_igcondapet}.

\noindent\textbf{Reason for the choices of different 3-tuples:} Since the attention module treats every pixel (token) as a key or query, the cost of computing attention grows quadratically with feature map size \cite{dosovitskiy2020image}. For an input of size $64\times 64$, the attention layer sees $64 \times 64 = 4096$ spatial tokens. This means that the self-attention forms a similarity matrix of size $4096^2 \approx $ 16.7 M elements to attend to every other token, which is computationally very expensive during training. Dropping down one level of UNet to $32\times 32$ cuts down the token to 1024 and the number of matrix entries to about 1.0 M, i.e., a 16$\times$ reduction in memory and FLOPs as compared to the level with first resolution level of size $64\times 64$. As a result, we only ablated over 3-tuples (000), (001), and (011), i.e., no attention was employed in the first level of the network. Additionally, the early levels of UNet mainly learns low-level local edge features and textures, which are already efficiently captured by convolutions. Moreover, the skip-connection from the first level to later level injects these details into the decoder so the model still sees the original fine-scale information even without attention there. For anomaly detection, the key benefit of attention is modeling longer-range, cross-organ context. Coarser levels ($16\times16$ or $32\times32$) are better suited for that because each token already represents a larger receptive field.

Furthermore, through empirical ablation experiments, we noted that the (001) and (011) variants, which insert attention only at $32\times32$ and/or $16 \times 16$, delivered similar or better performance on the chosen metrics than a variant with attention at all three levels, while using markedly less GPU memory \cite{ahamed2025weaklysupervisedpetanomalydetection}.

\subsection{Implicit-guidance}
\label{subsec:implicit_guidance}
Under the score-matching formulation of diffusion models \cite{song2019generative,hyvarinen2005estimation}, the score function is given by $\bfs_\theta(\bfx_t, t)=\nabla_{\bfx_t} \text{log}~p_\theta(\bfx_t, t)$, which represents the gradient of the log-likelihood with respect to $\bfx_t$. For a class-conditional diffusion model, a separate classifier $p_{\zeta}(c \mid \bfx_t)$ parametrized by $\zeta$ is used to bias the generative process towards samples of specific class during the generation process \cite{classifier_guidance_paper}. The classifier guidance modifies the score function by adding the gradient of the classifier's log-probability, $\Tilde{\bfs}_\theta(\bfx_t, c, t) =  \bfs_\theta(\bfx_t, c, t) ~+~ w \nabla_{\bfx_t} \text{log}~p_\zeta(c \mid \bfx_t)$, where $w$ is the guidance scale controlling the influence of classifier. The modified score $\Tilde{\bfs}_\theta(\bfx_t, c, t)$ is then used when sampling from the diffusion model, which has the effect of up-weighting the probability of data for which the classifier $\text{log}~p_\zeta(c \mid \bfx_t)$ assigns high likelihood to the correct label. This method, however, has several drawbacks: (i) it requires the training of a separate classifier alongside the diffusion model; (ii) the classifier adds additional computational overhead during sampling; and (iii) high guidance scale $w$ might improve sample quality but lead to mode collapse \cite{patel2023bridging}. Hence, in this work, we exploit implicit-guidance or \textit{classifier-free} guidance which removes the dependence on a separate classifier. 

In implicit-guidance \cite{classifier_free_guidance_paper}, the class-conditional embedding in the denoising UNet is leveraged to guide the model generation process. The denoising model predicts the noise $\boldsymbol{\epsilon}_\theta(\bfx_{t}, c, t)$ conditioned on class $c$. The key idea here is to compute an implicit ``score'' without explicitly using $\nabla_{\bfx_t} \text{log}~p_\zeta(c \mid \bfx_t)$. In implicit-guidance, the diffusion model is trained under dual objectives. We trained an unconditional denoising model to learn the unconditional distribution $p_\theta(\bfx_t, t)$ together with a conditional distribution $p_\theta(\bfx_t, c, t)$. A single denoising network $\boldsymbol{\epsilon}_\theta(\bfx_t, c, t)$ was used to parametrize both models, where the unconditional model was defined as $\boldsymbol{\epsilon}_\theta(\bfx_t, t) = \boldsymbol{\epsilon}_\theta(\bfx_t, c=0, t)$. 

The class labels for the healthy samples (images without lesions) and unhealthy samples (images with lesions)  were labeled as $c=1$ and $c=2$, respectively. The unconditional and conditional models were jointly trained by randomly setting $c=0$ to the unconditional class identifier with a probability of $p_\text{uncond} = 15\%$. The sampling was then performed using an updated estimate $\boldsymbol{\Tilde{\epsilon}}_\theta(\bfx_t, c, t)$ obtained by computing the linear combination of conditional and unconditional estimates,
\begin{equation}
    \label{eq:implicit_guidance}
    \boldsymbol{\Tilde{\epsilon}}_\theta(\bfx_t, c, t) = \boldsymbol{\epsilon}_\theta(\bfx_t, 0, t) + w \cdot \Big(\boldsymbol{\epsilon}_\theta(\bfx_t, c, t) - \boldsymbol{\epsilon}_\theta(\bfx_t, 0, t)\Big),
\end{equation}
where $c \in \{1,2\}$ and $w$ represents the guidance scale. The advantages of implicit-guidance over classifier guidance are as follows: (i) since the class-conditioning is integrated directly within the denoising network, both training and inference become more efficient; (ii) avoids overfitting to the classifier which might introduce bias or limitations in representing class information; and (iii) by directly conditioning on the class $c$ during the generative process, implicit guidance can produce samples that better align with the desired class, avoiding artifacts introduced by imperfect classifier gradients \cite{classifier_free_guidance_paper}.  

\subsection{Counterfactual generation and anomaly detection}
\label{subsec:counterfactual_generation_and_anomaly_detection}
Counterfactual generation in medical anomaly detection involves creating hypothetical scenarios to better understand and identify anomalies in medical data \cite{pearl2013structural, kenny2021generating}. In our work, counterfactual generation facilitates an ``unhealthy-to-healthy'' domain translation by generating a synthetic, healthy version of an input image, enabling the detection of anomalies through the calculated differences \cite{fontanella2024diffusion}. This method leverages minimal intervention in the generative process, ensuring the preservation of normal anatomical structures while highlighting pathological deviations.

\begin{algorithm}[h]
\caption{Anomaly detection using IgCONDA-PET}
\label{alg:icd}
\begin{algorithmic}[1]
\Require
    trained diffusion model $\boldsymbol{\epsilon}_\theta$ with 3-tuple attention-variant $(k_1k_2k_3)$; guidance scale $w$; number of iterations $D$;  
    input unhealthy image $\bfx_0$; class condition $c$
\Statex \textbf{Recovering unconditional latent space (encoding)}
\For{$t = 0$ \textbf{to} $D$}
    \State $\displaystyle 
        \hat{\bfx}_{t+1} \gets 
            \sqrt{\bar{\alpha}_{t+1}}
            \Bigg(\frac{\bfx_t - \sqrt{1-\bar{\alpha}_t}\,\epsilon_\theta(\bfx_t,0,t)}
                     {\sqrt{\bar{\alpha}_t}}\Bigg) + \sqrt{1 - \bar{\alpha}_{t+1}}\,\boldsymbol{\epsilon}_\theta(\bfx_t,0,t)$
\EndFor
\Statex \textbf{Counterfactual generation (decoding)}
\For{$t = D$ \textbf{to} $0$}
    \State $\boldsymbol{\epsilon} \gets w \boldsymbol{\epsilon}_\theta(\bfx_t,c,t) + (1-w)\boldsymbol{\epsilon}_\theta(\bfx_t,0,t)$
    \State $\displaystyle
        \hat{\bfx}_{t-1} \gets
            \sqrt{\bar{\alpha}_{t-1}}
            \Bigg(
                \frac{\bfx_t - \sqrt{1-\bar{\alpha}_t}\,\boldsymbol{\epsilon}}{\sqrt{\bar{\alpha}_t}}
            \Bigg) + \sqrt{1 - \bar{\alpha}_{t-1}}\,\boldsymbol{\epsilon}$
\EndFor
\State \Return $\displaystyle
        \text{heatmap} = \bigl|\bfx_0 - \hat{\bfx}_0\bigr|$
\end{algorithmic}
\end{algorithm}

During inference, we first set a noise level $D \in \{1, ..., T\}$ and a guidance scale $w$. Starting with an unhealthy input image $\bfx_0$ (with $c=2$), we perform noise encoding to obtain a latent image $\bfx_D$ by iteratively applying $\boldsymbol{\epsilon}_\theta(\bfx_t, 0, t)$ using the reverse of \Cref{eq:ddim_sampling}. During denoising, a copy of generated $x_D$ was fed into the unconditional model $\boldsymbol{\epsilon}_\theta(\bfx_t, 0, t)$ and the model with healthy conditioning $\boldsymbol{\epsilon}_\theta(\bfx_t, c=1, t)$ to obtain the updated estimate of noise via \Cref{eq:implicit_guidance}, followed by denoising for $D$ steps via the schedule in \Cref{eq:ddim_sampling} to generate the counterfactual or the corresponding pseudo-healthy image $\hat{\bfx}_0 (c=1)$. The anomaly map was computed using the absolute difference between the original unhealthy input and the generated healthy counterfactual using,
\begin{equation}
    \label{eq:anomaly_map}
    \text{AM}(\bfx_0(c=2)) = \Big|\bfx_0(c=2) - \hat{\bfx}_0(c=1)\Big|,
\end{equation}
which can be used to obtain the location of anomalies. A schematic of the method has been shown in \Cref{fig:method_scheme} and an algorithm for anomaly detection is summarized in \Cref{alg:icd}.

\section{Experiments}
\label{sec:experiments}
\subsection{Datasets and preprocessing}
\label{subsec:datasets_and_preprocessing}
In this work, we used a large, diverse, multi-institutional, multi-cancer, and multi-tracer PET datasets with a total of \textbf{2652 cases}. These scans came from six retrospective cohorts, consisting of local and public datasets: 
\begin{enumerate}
\item \textbf{AutoPET}: This public dataset with 1611 cases came from the AutoPET-III challenge 2024 hosted at MICCAI 2024 \cite{autopet_paper}. These scans consisted of two sub-cohorts: (a) 1014 $^{18}$F-FDG PET scans from 900 patients spanning various cancer types such as lymphoma (146 scans), lung cancer (169 scans), melanoma (191 scans) as well as negative control patients (508 scans), and (b) 597 PSMA PET scans (369 $^{18}$F-PSMA and 228 $^{68}$Ga-PSMA) from 378 patients with prostate cancer. The FDG data was acquired at University Hospital T{\"u}bingen, Germany, while the PSMA data was acquired at LMU Hospital, LMU Munich, Germany. 
\item \textbf{HECKTOR}: This public dataset with 524 FDG-PET cases came from the HECKTOR 2022 challenge hosted at MICCAI 2022 \cite{hecktor_paper}. These consisted of 524 patients with head \& neck cancer from 7 different centers across North America and Europe: (i) University of Montreal Hospital Center, Montreal, Canada (56 scans), (ii) Sherbrooke University Hospital Center, Sherbrooke, Canada (72 scans) (iii) Jewish General Hospital, Montreal, Canada (55 scans), (iv) Maisonneuve-Rosemont Hospital, Montreal, Canada (18 scans), (v) MD Anderson Cancer Center, Houston, Texas, USA (198 scans), (vi) Poitiers University Hospital, France (72 scans), and (vii) Vaudois University Hospital, Switzerland (53 scans).      
\item \textbf{DLBCL-BCCV}: This private dataset consisted of 107 $^{18}$F-FDG PET scans from 79 patients with diffuse large B-cell lymphoma (DLBCL) from BC Cancer, Vancouver (BCCV), Canada \cite{clinical_metrics_paper}. 
\item \textbf{PMBCL-BCCV}: This private dataset consisted of 139 $^{18}$F-FDG PET scans from 69 patients with primary mediastinal B-cell lymphoma (PMBCL) from BCCV \cite{clinical_metrics_paper}. 
\item \textbf{DLBCL-SMHS}: This private dataset consisted of 220 $^{18}$F-FDG PET scans from 219 patients with DLBCL from St. Mary's Hospital, Seoul (SMHS), South Korea \cite{clinical_metrics_paper}.
\item \textbf{STS}: This public dataset consisted of 51 $^{18}$F-FDG PET scans from patients with soft-tissue sarcoma \cite{vallieres2015radiomics}.
\end{enumerate}
While the cohorts 1-5 were used for both model developed and (internal) testing, the cohort 6 was used solely for external testing. Additionally, 10 images (2 each from datasets 1-5) were randomly sampled and set aside solely for performing hyperparameter sensitivity experiments (see, \Cref{subsec:sensitivity_to_inference_hyperparameteres}). The training, validation and test set splits were stratified at the patient level to avoid multiple images from the same patient being shared among training and validation/test sets. The training, validation and test splits consisted of 2210, 149, and 283, respectively. The test sets (internal or external) excluded any images from negative control patients (originating from cohort 1). The ethical statements about these datasets can be found in \Cref{ethical_statements}. 

All these datasets consisted of densely annotated manual segmentations by physicians (i.e. at the pixel level). Since the AutoPET FDG cohort was the largest cohort used in this work, we resampled the images from all other cohorts to the voxel spacing of the AutoPET FDG cohort (2.0 mm, 2.0 mm, 3.0 mm). The resampling was performed using bilinear interpolation for PET scans and nearest-neighbor interpolation for densely annotated masks. All scans were centrally cropped using a 3D bounding box of size $192 \times 192 \times 288$ and then downsampled to $64 \times 64 \times 96$ ($\times3$ downsampling). The dense voxel-level annotations were used to define the image-level labels as healthy ($c=1$) or unhealthy ($c=2$) for the axial slices for each scan. The fraction of unhealthy slices across the six datasets (excluding the negative control patients from dataset 1) were 24.4\%, 12.8\%, 8.4\%, 8.0\%, 24.4\%, and 14.5\% respectively showing the diversity of our datasets.

{No cross-center harmonization (such as ComBat \cite{leithner2022impact} or style-transfer CNNs \cite{liu2023style}) was employed in our work, which is in line with the common practice of recent deep-learning based studies on PET lesion segmentation \cite{andrearczyk2023head,gatidis2024results}. Two main considerations motivated this decision: (i) Harmonization mappings are center-specific and can make a model underperform when deployed on scanners or reconstruction kernels unseen during training. By performing no cross-center standardization, we expose our network to the full spectrum of inter-scanner variability, encouraging the self-attention layers to learn scanner-invariant features rather than relying on brittle pre-processing steps; (ii) Global histogram alignment dulls the fine SUV gradients that reveal small or low-uptake lesions; preserving raw intensities and thereby preserving lesion contrast. Instead, we adopt the lightweight pre-processing used by most recent deep-learning works \cite{andrearczyk2023head,gatidis2024results}, namely converting PET intensity values to SUV, resampling to a common grid, standard translation, rotation, scaling-based data augmentation during training, etc. Training the diffusion network on these minimally processed volumes exposes it to genuine scanner variability and lets the multi-scale self-attention learn scanner-invariant features.

\subsection{Training protocol and implementation}
\label{subsec:training_protocol}
The parameters $\theta$ and $\phi$ of the 2D denoising diffusion UNet and the embedding layers for class conditioning respectively were trained using Adam optimizer with a learning rate of $10^{-5}$. The model with the lowest validation MSE loss over 1000 epochs was used for test set evaluation. During training, we set $T = 1000$ and batch size = 64. Data augmentation strategies were employed during training to prevent the model from over-fitting to scanner- or patient-specific artifacts. These augmentations included (i) 2D translations in the range (0, 10) pixels along the two dimensions on the axial plane, (ii) axial rotations by angle in range $(-\pi/12, \pi/12)$, (iii) random scaling by a factor of 1.1 along the two dimensions, (iv) 2D elastic deformations using a Gaussian kernel with standard deviation and offsets on the grid uniformly sampled from (0, 1), (v) Gamma correction with $\gamma \in (0.7, 1.5)$. All experiments were performed on a Microsoft Azure virtual machine with NVIDIA Tesla V100 GPUs with a collective GPU memory of 64 GiB and 448 GiB RAM. All implementations were done in Python 3.12.4, PyTorch 2.4.0, and MONAI 1.3.2 \cite{monai_generative_paper}. The code is publicly available at: \url{https://github.com/ahxmeds/IgCONDA-PET.git}.

\subsection{Baselines}
\label{subsec:benchmark}
The performance of our best performing model, IgCONDA-PET(0,1,1), were compared against several deep learning based weakly-supervised or unsupervised methods as well as some traditional methods like thresholding. These are summarized below:

\begin{enumerate}
    \item \textbf{41\% SUV$_\text{max}$ thresholding:} This is a common automated method for segmenting lesions in PET images, utilizing the SUV$_\text{max}$ value in the whole-body images. The technique involves defining a threshold, commonly, 41\% of the SUV$_\text{max}$ value. The voxels with values above/below this threshold are labeled as unhealthy/healthy. Although computationally simple and easy to implement, this method is highly prone to errors due to the presence of physiological high-uptake regions like brain, bladder, kidneys, heart, etc, which also get segmented as lesions via this method \cite{thresholding_41percent}. 
    
    \item \textbf{ResNet-18 classifier with class-activation map (CAM) explanation:} This method was adapted a previous study \cite{ahamed2023slice} that utilized a deep classifier with ResNet-18 backbone to classify PET axial slices into slices containing and not containing lesions. The ResNet-18 backbone consisted of ImageNet-pretrained weights that were fine-tuned on PET datasets. Additionally, we utilized CAM-based explanation originating from the feature maps of the last convolutional layer to provide interpretable visual explanation behind the decisions made by the classifier, which were visualized as heatmaps in the lesion regions for unhealthy slices. We tried various popular CAM techniques such as GradCAM, GradCAM++, HiResCAM, ScoreCAM, AblationCAM, XGradCAM, EigenCAM, and FullGradCAM \cite{poppi2021revisiting,draelos2020use, ramaswamy2020ablation,muhammad2020eigen,srinivas2019full}. In this work, we only report the performance of HiResCAM \cite{draelos2020use} since it had the best performance among all other CAM methods.   
    
    \item \textbf{f-AnoGAN:} This is an improved and efficient version of AnoGAN, an anomaly detection framework based on Generative Adversarial Networks (GANs) based on \cite{fanogan_paper}. 
    This method consists of three modules, an encoder, a generator and a discriminator. The encoder maps the data to the latent space of the generator. The generator learns to reconstruct normal data generating realistic outputs from a latent space representation. The discriminator module evaluates the quality of the generated samples distinguishing real data and fake (generated) outputs during training. Finally, the input data is compared with its reconstruction from the trained generator. Larger reconstruction errors indicate higher likelihoods of anomalies, thereby, highlighting lesions.

    \item \textbf{VT-ADL:} Vision Transformer for Anomaly Detection and Localization (VT-ADL) is an unsupervised reconstruction-based framework that combines the global‐context modeling power of a Vision Transformer (ViT) encoder with a lightweight convolutional decoder \cite{mishra2021vt}.  
    During training, only healthy PET slices are shown to the network; the ViT encoder (initialized with ImageNet weights) extracts patch tokens, which the decoder upsamples back to the image grid. The model is optimized with a voxel-wise mean-squared-error loss between the input and its reconstruction. At inference, the absolute reconstruction error is interpreted as an anomaly map: voxels (or patches) whose intensities cannot be faithfully reproduced by the healthy-trained auto-encoder receive higher scores and are flagged as potential lesions. Compared with purely CNN-based autoencoders, the transformer backbone allows VT-ADL to capture long-range dependencies, yielding sharper localization of irregular uptake patterns while keeping the network lightweight and fast to train.

    \item \textbf{DPM with classifier guidance (DPM+CG):} This diffusion method integrates a classifier's predictions to modify the generative path during the reverse diffusion stages, essentially using the classifier's output to guide the synthesis of images by reinforcing features that align with specific class attributes \cite{classifier_guidance_paper,diff_model_med_anomaly_detection}. In this work, used the same diffusion network as IgCONDA-PET (for fair comparison) with the (011) attention variant but without the class-conditioning input. Furthermore, we used a classifier with 3 levels (also with attention mechanism (011)) consisting of 32, 64, 64 channels in 3 layers. The diffusion model and classifier were trained independently. During inference, the classifier was used to modify the denoising path by adding a scaled gradient of log probability (from the classifier), as explained in \Cref{subsec:implicit_guidance}, to generate a healthy counterfactual. We used the optimal denoising steps $D=200$ and optimal guidance scale $w=6.0$ which were obtained on a separate validation set of 10 patients (from the internal cohorts). The anomaly map was similarly generated by computing the absolute difference between the unhealthy input and the generated unhealthy counterfactual. 

\end{enumerate}

For fair comparison, all the baselines were developed on $64 \times 64$ images, except VT-ADL (pretrained ViT backbone in VT-ADL required $224\times224$ inputs, the anomaly maps produced were later downsampled to $64\times64$), although most other training and inference hyperparameters were adapted from the original works and fine-tuned wherever necessary to stabilize training.

\subsection{Evaluation metrics}
\label{subsec:evaluation_metrics}
Our anomaly detection methods generated anomaly maps with values in range [0,1]. Hence, we employed anomaly map binarization to compute metrics at different thresholds $\tau$. We evaluated the anomaly detection performance using various metrics: (i) Optimal Dice similarity coefficient (DSC) or $\lceil \text{DSC} \rceil$ \cite{what_is_healthy}; (ii) 95\%tile Hausdorff distance (HD95) in pixels; (iii) Area under the precision-recall curve (AUPRC); (iv) Lesion detection sensitivity. 

$\lceil \text{DSC} \rceil$ and HD95 are slice-level metrics, detection sensitivity is computed at the level of lesion (2D lesion on axial slices), while AUPRC is a pixel-level metric. $\lceil \text{DSC} \rceil$ and HD95 provide insights into overall segmentation accuracy and boundary precision at the level of individual slices, assessing how well the model captures the overall shape and structure of anomalies. Lesion detection sensitivity assesses the model's ability to identify entire lesions as coherent entities, which is crucial for detecting clinically significant abnormalities. By evaluating at the lesion level, this metric accounts for real-world diagnostic requirements \cite{clinical_metrics_paper}. The AUPRC evaluates the model's ability to correctly classify anomalous pixels, focusing on more fine-grained accuracy. This is especially important for detecting subtle anomalies that may be missed at coarser levels of analysis, thereby ensuring that the model is sensitive to variations in pixel-level abnormality. By combining these metrics, the analyses captures the performance from different perspectives—global accuracy (slice-level), clinical relevance (lesion-level) and detailed precision (pixel-level). This multi-faceted approach ensures that the anomaly detection model is robust, accurate, and clinically applicable across a range of use cases. We describe these metrics in detail next. 

Let the Dice similarity coefficient (DSC) between a ground truth binary mask $\bfg$ and a predicted binarized  anomaly map $\bfp(\tau)$ (using threshold $\tau$) be given by, 
\begin{equation}
    \text{DSC}(\bfg, \bfp(\tau)) = \frac{2 |\bfg \cap \bfp(\tau)|}{|\bfg| + |\bfp(\tau)|}.
\end{equation}
We compute the optimal threshold $\tau^*$ for binarization by sweeping over thresholds in the range 0.1-0.9 and choosing the best $\tau$ that maximizes DSC, 
\begin{equation}
    \tau^* = \argmax_{\tau} \text{DSC}(\bfg, \bfp(\tau))
\end{equation}
Hence, the optimal DSC metric was obtained using, 
\begin{equation}
   \lceil \text{DSC}  \rceil(\bfg, \bfp) = \text{DSC}(\bfg, \bfp(\tau^*)) 
\end{equation}
Secondly, using the same $\tau^*$, we compute HD95 as follows, 
\begin{equation}
\text{HD95}(\bfg, \bfp) = \max\{\text{Percentile}_{95} (d(\bfg, \bfp(\tau^*))), \text{Percentile}_{95} (d(\bfp(\tau^*), \bfg))\}
\end{equation}
where $d(\bfg, \bfp(\tau))$ represents the set of all distances from each point on the boundary of ground truth lesions in $\bfg$ to its nearest point on the boundary of predicted lesions in $\bfp(\tau)$, and $d(\bfp(\tau), \bfg)$ represents the set of all distances from each point on the boundary of predicted lesions in $\bfp(\tau)$ to its nearest point on the boundary of lesions in $\bfg$. We used the 95\%tile value to make the metric robust to  noise or outliers in either mask because it ignores the top 5\% of distances. 

We evaluated the detection sensitivity at the lesion level, adapted from \cite{clinical_metrics_paper}, where it was referred to as the \textit{lesion detection sensitivity under detection criterion 2}. Let the set of (disconnected) lesions contained in ground truth mask $\bfg$ be $\{\bfg_1, \bfg_2, \ldots, \bfg_L\}$ and the set of lesions contained in predicted mask $\bfp(\tau^*)$ be $\{\bfp_1(\tau^*), \bfp_2(\tau^*), \ldots, \bfp_{L^\prime}(\tau^*)\}$, where $L$ and $L^\prime$ denote the number of lesions in the ground truth and predicted binarized anomaly map, respectively. For computing detection sensitivity, all the predicted lesion were first matched to their corresponding ground truth lesions by maximizing the Intersection-Over-Union (IoU) between each predicted and ground truth lesions pair. For a predicted lesion $\bfp_{l^\prime}(\tau^*)$ matched to a ground truth lesion $\bfg_l$, $\bfp_{l^\prime}(\tau^*)$ was labeled as true positive if,
\begin{equation}
    \text{IoU}(\bfg_l, \bfp_{l^\prime}(\tau^*)) = \frac{|\bfg_l \cap \bfp_{l^\prime}(\tau^*)|}{|\bfg_l \cup \bfp_{l^\prime}(\tau^*)|} \geq 0.5.
\end{equation}
From this notion of true positive, we computed the detection sensitivity by computing the ratio of true positive lesions in $\bfp$ to the total number lesions in $\bfg$. 

Finally, we also evaluated AUPRC (at the pixel level) between a ground truth mask $\bfg$ and the predicted (unthresholded) anomaly map $\bfp$. Let the set of pixels in $\bfg$ and $\bfp$ be denoted by $\{g^{(1)}, g^{(2)}, \ldots, g^{(N)}\}$ and $\{p^{(1)}, p^{(2)}, \ldots, p^{(N)}\}$ respectively, where $N$ denotes the total number of pixels on the slice. Here, we assume that the two sets have been sorted in the descending order of the pixel values in $p$. The AUPRC is then computed as a discrete sum, 
\begin{equation}
\text{AUPRC}(\bfg, \bfp) = \sum_{i=1}^N \Big(\frac{\text{TP}_i}{P} - \frac{\text{TP}_{i-1}}{P} \Big) \cdot \frac{\text{TP}_{i}}{i},
\end{equation}
where $\text{TP}_i = \sum_{k=1}^i g^{(k)}$ represents the number of true positive pixels among the top $i$ sorted predictions, $P =  \sum_{k=1}^N g^{(k)}$ represents the total number of true positives, $\text{TP}_i/P$ represents the recall value at rank $i$, and $\text{TP}_i/i$ represents the precision value at rank $i$ (TP$_0 = 0$ by convention). Here, the sorted values $\{p^{(1)}, p^{(2)}, \ldots, p^{(N)}\}$ are treated as different thresholds for AUPRC analyses.

\section{Results}
\label{sec:results}

\subsection{Test set performance and benchmarking}
\label{subsec:test_set_performance_and_benchmarking}

\begin{table}[H]
\centering
\resizebox{\textwidth}{!}{%
\begin{tabular}{@{}ccccccc@{}}
\toprule
\multirow{3}{*}{Methods} & \multicolumn{6}{c}{$\lceil \text{DSC} \rceil (\%) (\uparrow)$}                                                                          \\ \cmidrule(l){2-7} 
                         & \multicolumn{5}{c}{Internal testing}                                          & External testing \\ \cmidrule(l){2-7} 
                         & AutoPET & HECKTOR         & DLBCL-BCCV  & PMBCL-BCCV  & DLBCL-SMHS  & STS              \\ \midrule
Thresholding$^*$ \cite{thresholding_41percent}             & 4.6 $\pm$ 12.1   & 31.5 $\pm$ 33.3          & 22.1 $\pm$ 28.7 & 1.7 $\pm$ 4.6   & 13.9 $\pm$ 21.6 & 16.3 $\pm$ 23.3    \\
Classifier$^\ddagger$  + HiResCAM$^*$ \cite{ahamed2023slice}    & 16.6 $\pm$ 17.5  & 21.8 $\pm$ 18.9          & 25.9 $\pm$ 19.9 & 14.2 $\pm$ 21.8 & 20.6 $\pm$ 18.6 & 28.9 $\pm$ 23.8      \\
f-AnoGAN$^\dagger$ \cite{fanogan_paper}                  & 44.2 $\pm$ 24.5  & \textbf{57.0 $\pm$ 25.3} & 42.7 $\pm$ 19.9 & 46.9 $\pm$ 29.7 & 51.0 $\pm$ 20.7 & 48.9 $\pm$ 22.7      \\
VT-ADL$^\dagger$ \cite{mishra2021vt}                  & 30.0 $\pm$ 28.1  & 46.9 $\pm$ 28.2 &  42.9 $\pm$ 34.6 & 21.0 $\pm$ 26.4 & 49.3 $\pm$ 33.3 & 47.9 $\pm$ 33.3      \\
DPM + CG$^\ddagger$ \cite{classifier_guidance_paper}                 & 38.8 $\pm$ 21.2  & 27.2 $\pm$ 19.4          & 32.3 $\pm$ 15.5 & 35.5 $\pm$ 23.0 & 42.1 $\pm$ 18.1 & 26.8 $\pm$ 16.8      \\ \midrule
IgCONDA-PET (000)$^\ddagger$  & 45.9 $\pm$ 27.9          & 49.0 $\pm$ 28.6 & 49.9 $\pm$ 26.3          & 32.5 $\pm$ 29.5          & 56.6 $\pm$ 24.8          & 49.6 $\pm$ 28.6          \\
IgCONDA-PET (001)$^\ddagger$  & 48.7 $\pm$ 26.7          & 50.8 $\pm$ 29.4 & 51.7 $\pm$ 24.3          & \textbf{56.3 $\pm$ 30.3} & 57.6 $\pm$ 24.5          & 51.3 $\pm$ 28.6          \\
IgCONDA-PET (011)$^\ddagger$  & \textbf{49.4 $\pm$ 26.7} & 54.7 $\pm$ 27.8 & \textbf{53.0 $\pm$ 25.5} & 56.0 $\pm$ 29.6          & \textbf{60.3 $\pm$ 24.4} & \textbf{51.0 $\pm$ 30.5} \\ \bottomrule
\end{tabular}%
}
\caption{Quantitative comparison using the \textbf{$\lceil \text{DSC} \rceil$} metric (higher is better) between different anomaly detection methods on the test sets. Performances of the top models in each column has been shown in bold. $\pm$indicates standard deviation across all unhealthy slices. Here, $*$: ``not trained'', ${\dagger}$: ``trained on only healthy data'', and ${\ddagger}$: ``trained on both healthy and unhealthy data''.}
\label{tab:dsc_metric}
\end{table}

\begin{table}[h]
\centering
\resizebox{\textwidth}{!}{%
\begin{tabular}{@{}ccccccc@{}}
\toprule
\multirow{3}{*}{Methods} & \multicolumn{6}{c}{HD95 (pixel) $(\downarrow)$}                                                                               \\ \cmidrule(l){2-7} 
                         & \multicolumn{5}{c}{Internal testing}                                                & External testing \\ \cmidrule(l){2-7} 
                         & AutoPET  & HECKTOR & DLBCL-BCCV           & PMBCL-BCCV    & DLBCL-SMHS    & STS              \\ \midrule
Classifier$^\ddagger$ + HiResCAM$^*$ \cite{ahamed2023slice}    & 16.3 $\pm$ 11.6 & 9.3 $\pm$ 7.8  & 12.3 $\pm$ 11.3        & 20.7 $\pm$ 13.6 & 16.4 $\pm$ 11.7 & 20.9 $\pm$ 15.4    \\
f-AnoGAN$^\dagger$ \cite{fanogan_paper}                 & 12.6 $\pm$ 11.8 & 7.9 $\pm$ 6.7  & 10.9 $\pm$ 10.8        & 10.9 $\pm$ 12.1 & 8.4 $\pm$ 9.6   & 16.6 $\pm$ 11.7    \\
VT-ADL$^\dagger$ \cite{mishra2021vt}                  & 22.2 $\pm$ 16.6  & 8.1 $\pm$ 11.3 &  15.5 $\pm$ 16.7 & 29.1 $\pm$ 15.8 & 15.4 $\pm$ 16.3 & 18.1 $\pm$ 18.6      \\
DPM + CG$^\ddagger$ \cite{classifier_guidance_paper}                 & 14.8 $\pm$ 10.9 & 12.8 $\pm$ 7.1 & 18.1 $\pm$ 10.7        & 14.6 $\pm$ 11.8 & 12.5 $\pm$ 9.5  & 24.5 $\pm$ 12.0    \\ \midrule
IgCONDA-PET (000)$^\ddagger$        & 12.8 $\pm$ 12.6 & 8.8 $\pm$ 9.6  & 10.8 $\pm$ 11.1        & 18.8 $\pm$ 17.1 & 7.9 $\pm$ 9.0   & 16.1 $\pm$ 16.3    \\
IgCONDA-PET (001)$^\ddagger$        & 11.1 $\pm$ 10.6 & 6.6 $\pm$ 6.1  & \textbf{8.8 $\pm$ 9.8} & 8.9 $\pm$ 9.9   & 8.2 $\pm$ 9.3   & 15.6 $\pm$ 14.8    \\
IgCONDA-PET (011)$^\ddagger$ & \textbf{10.6 $\pm$ 10.3} & \textbf{5.5 $\pm$ 5.8} & 9.3 $\pm$ 10.0 & \textbf{7.4 $\pm$ 7.9} & \textbf{6.9 $\pm$ 8.4} & \textbf{15.4 $\pm$ 16.1} \\ \bottomrule
\end{tabular}%
}
\caption{Quantitative comparison using the \textbf{HD95} metric (lower is better) in pixels between different anomaly detection methods on the test sets. Performances of the top models in each column has been shown in bold.}
\label{tab:hd95_metric}
\end{table}

\begin{figure}[H]
\centering
\includegraphics[width=\textwidth]{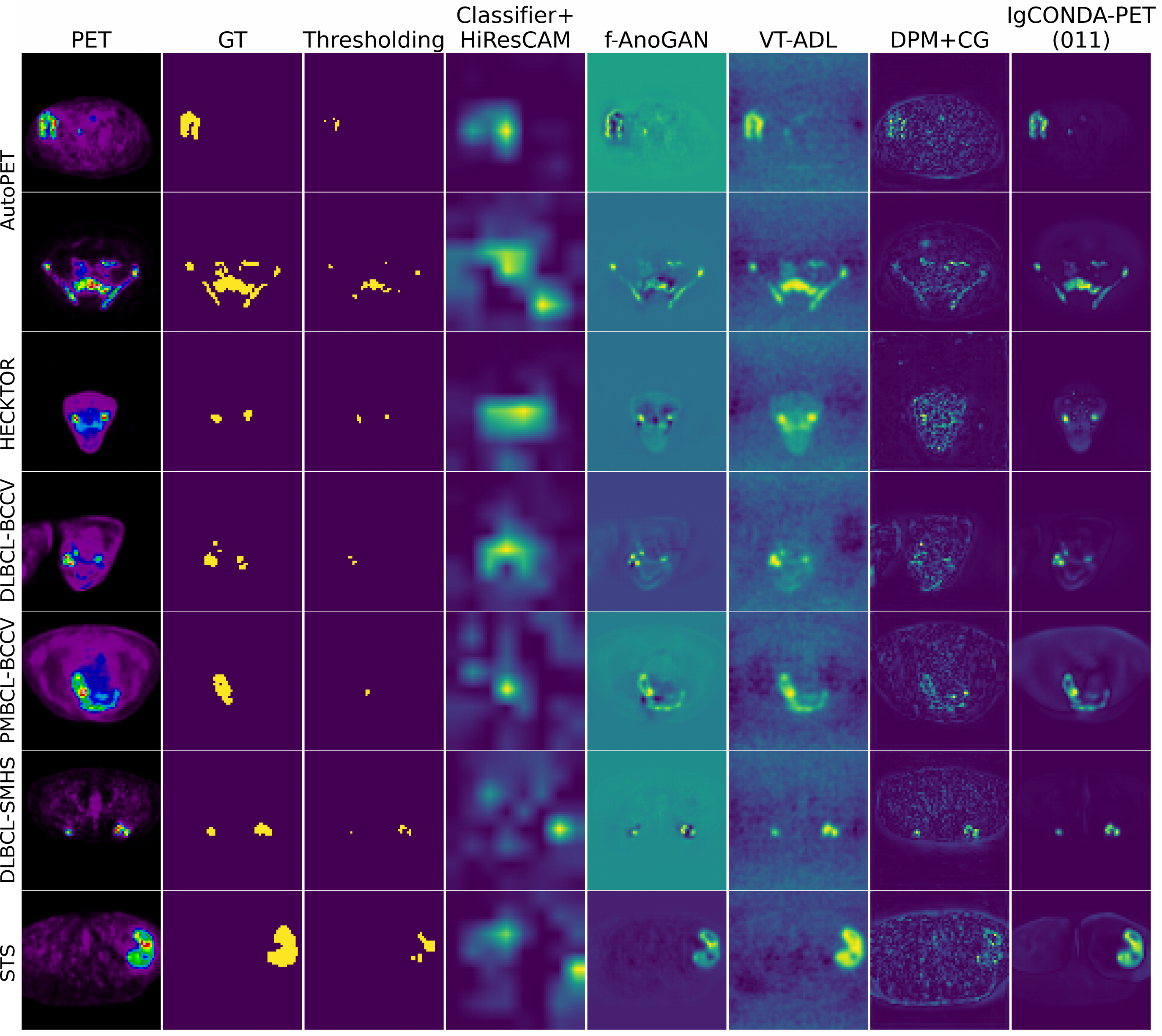}
\caption{Qualitative comparison between the anomaly maps generated by different methods such as 41\% SUV$_\text{max}$ Thresholding, Classifier+HiResCAM, f-AnoGAN, VT-ADL, DPM+CG, and our method IgCONDA-PET (011) on PET slices. GT represents the physician’s dense ground truth. Our approach delivers the most precise lesion localization and keeps non-pathological regions virtually free of spurious activations (unlike other baseline methods like f-AnoGAN or VT-ADL). Here, for IgCONDA-PET (011), the anomaly map was generated using the inference hyperparameters $D = 400$ and $w = 3.0$.}
\label{fig:compare_different_methods_visualization}
\end{figure}

\begin{table}[h]
\centering
\resizebox{\textwidth}{!}{%
\begin{tabular}{@{}ccccccc@{}}
\toprule
\multirow{3}{*}{Methods} & \multicolumn{6}{c}{AUPRC (\%) $(\uparrow)$}                                                                                 \\ \cmidrule(l){2-7} 
                         & \multicolumn{5}{c}{Internal testing}                                                   & External testing \\ \cmidrule(l){2-7} 
                         & AutoPET  & HECKTOR           & DLBCL-BCCV    & PMBCL-BCCV    & DLBCL-SMHS    & STS              \\ \midrule
Classifier$^\ddagger$ + HiResCAM$^*$ \cite{ahamed2023slice}   & 11.2 $\pm$ 17.1 & 15.4 $\pm$ 18.9          & 17.7 $\pm$ 19.0 & 12.0 $\pm$ 23.0 & 14.3 $\pm$ 17.6 & 24.6 $\pm$ 25.9    \\
f-AnoGAN$^\dagger$ \cite{fanogan_paper}                & 40.9 $\pm$ 28.0 & \textbf{53.8 $\pm$ 29.0} & 37.5 $\pm$ 22.3 & 44.1 $\pm$ 32.5 & 48.0 $\pm$ 23.3 & 46.0 $\pm$ 27.1    \\
VT-ADL$^\dagger$ \cite{mishra2021vt}                  & 28.1 $\pm$ 30.5  & 47.5 $\pm$ 33.1 &  43.8 $\pm$ 37.6 & 19.9 $\pm$ 29.2 & 50.6 $\pm$ 37.5 & 47.7 $\pm$ 38.0     \\
DPM + CG$^\ddagger$ \cite{classifier_guidance_paper}                & 31.2 $\pm$ 23.3 & 18.9 $\pm$ 20.3          & 23.1 $\pm$ 14.8 & 29.9 $\pm$ 24.5 & 36.5 $\pm$ 20.8 & 22.9 $\pm$ 16.9    \\ \midrule
IgCONDA-PET (000)$^\ddagger$        & 43.7 $\pm$ 31.8 & 46.0 $\pm$ 33.1          & 47.8 $\pm$ 31.2 & 30.9 $\pm$ 32.2 & 56.9 $\pm$ 29.3 & 49.6 $\pm$ 32.9    \\
IgCONDA-PET (001)$^\ddagger$ & 46.8 $\pm$ 31.4          & 48.4 $\pm$ 34.0 & 48.7 $\pm$ 29.7          & \textbf{55.7 $\pm$ 35.3} & 56.3 $\pm$ 29.6          & 48.0 $\pm$ 32.8          \\
IgCONDA-PET (011)$^\ddagger$ & \textbf{47.7 $\pm$ 31.5} & 52.0 $\pm$ 32.8 & \textbf{51.3 $\pm$ 30.7} & 54.4 $\pm$ 35.2          & \textbf{60.7 $\pm$ 29.4} & \textbf{50.7 $\pm$ 34.6} \\ \bottomrule
\end{tabular}%
}
\caption{Quantitative comparison using the \textbf{AUPRC} metric (higher is better) between different anomaly detection methods on the test sets. Performances of the top models in each column has been shown in bold.}
\label{tab:auprc_metric}
\end{table}

\begin{table}[h]
\centering
\resizebox{\textwidth}{!}{%
\begin{tabular}{@{}ccccccc@{}}
\toprule
\multirow{3}{*}{Methods} & \multicolumn{6}{c}{Detection sensitivity (\%) $(\uparrow)$}                                                               \\ \cmidrule(l){2-7} 
                         & \multicolumn{5}{c}{Internal testing}                                          & External testing \\ \cmidrule(l){2-7} 
                         & AutoPET  & HECKTOR  & DLBCL-BCCV    & PMBCL-BCCV    & DLBCL-SMHS    & STS              \\ \midrule
Thresholding$^*$ \cite{thresholding_41percent}             & 0.9 $\pm$ 7.5   & 24.6 $\pm$ 39.2 & 10.2 $\pm$ 26.2 & 0.0 $\pm$ 0.0   & 3.3 $\pm$ 17.2  & 7.4 $\pm$ 26.1     \\
Classifier$^\ddagger$ + HiResCAM$^*$ \cite{ahamed2023slice}    & 2.1 $\pm$ 13.5  & 2.1 $\pm$ 13.2  & 4.2 $\pm$ 20.1  & 7.8 $\pm$ 27.2  & 1.6 $\pm$ 11.0  & 9.4 $\pm$ 29.1     \\
f-AnoGAN$^\dagger$ \cite{fanogan_paper}                & 37.4 $\pm$ 37.9 & 53.1 $\pm$ 44.4 & 26.6 $\pm$ 37.8 & 42.2 $\pm$ 48.3 & 36.4 $\pm$ 39.7 & 30.9 $\pm$ 46.2    \\
VT-ADL$^\dagger$ \cite{mishra2021vt}                  &  21.8 $\pm$ 33.8  & 36.0 $\pm$ 43.4 &  43.4 $\pm$ 44.3 & 23.5 $\pm$ 42.8 & 43.9 $\pm$ 42.8 & 47.3 $\pm$ 49.9     \\
DPM + CG$^\ddagger$ \cite{classifier_guidance_paper}                & 31.4 $\pm$ 35.9 & 24.6 $\pm$ 38.2 & 21.0 $\pm$ 34.1 & 22.5 $\pm$ 41.6 & 25.1 $\pm$ 35.1 & 4.7 $\pm$ 21.1     \\ \midrule
IgCONDA-PET (000)$^\ddagger$        & 42.3 $\pm$ 38.8 & 51.0 $\pm$ 44.5 & 44.0 $\pm$ 43.8 & 42.2 $\pm$ 49.4 & 51.7 $\pm$ 41.3 & 44.5 $\pm$ 50.0    \\
IgCONDA-PET (001)$^\ddagger$        & 44.6 $\pm$ 39.1 & 52.2 $\pm$ 44.8 & 47.8 $\pm$ 43.7 & 56.7 $\pm$ 45.5 & 54.4 $\pm$ 42.2 & 45.6 $\pm$ 49.7    \\
IgCONDA-PET (011)$^\ddagger$ & \textbf{45.8 $\pm$ 39.2} & \textbf{54.3 $\pm$ 44.1} & \textbf{54.3 $\pm$ 42.5} & \textbf{58.8 $\pm$ 48.7} & \textbf{56.9 $\pm$ 41.3} & \textbf{48.3 $\pm$ 49.9} \\ \bottomrule
\end{tabular}%
}
\caption{Quantitative comparison using the lesion \textbf{detection sensitivity} metric (higher is better) between different anomaly detection methods on the test sets. Performances of the top models in each column has been shown in bold.}
\label{tab:sensitivity}
\end{table}

The quantitative performances for different methods have been presented in \Cref{tab:dsc_metric,tab:hd95_metric,tab:auprc_metric,tab:sensitivity} for the metrics $\lceil \text{DSC} \rceil$ (higher is better), HD95 (lower is better), AUPRC (higher is better), and lesion detection sensitivity (higher is better) over different internal and external test sets. From these tables, it can be seen that our method IgCONDA-PET (especially the (011) version) performs the best across all metrics on most of the test sets. Although, on the HECKTOR test set, f-AnoGAN outperformed IgCONDA-PET (011) by 2.3\% on $\lceil \text{DSC} \rceil$ and 1.8\% on AUPRC although performed worse by 2.4 pixels on HD95 and 1.2\% on detection sensitivity. 

As expected, the thresholding method often only highlights the hottest region within the lesions (see \Cref{fig:compare_different_methods_visualization}) leading to lower anomaly detection performance. Classifier with CAM explanation, although often highlight the regions around lesions, these proposed regions are not often tight enough with respect to the ground truth lesion boundaries, leading to worse performance. The anomaly maps generated via f-AnoGAN shows high intensity regions in the healthy regions of the unhealthy slices, often obscuring the highlighted anomaly. A similar behavior was observed for VT-ADL, which produces considerable spurious activations in the non-pathological regions, diminishing the overall anomaly detection performance.} DPM+CG method too generates anomaly maps with higher intensities in the healthy regions of the unhealthy slices which prevents it from successfully capturing the high-intensity anomalies on the slice.  

To clarify where IgCONDA-PET (011)’s advantage comes from, we systematically contrast it with targeted ablations and baseline networks, isolating the architectural elements that drive its superior performance:
\begin{enumerate}
    \item \textbf{Architectural factors behind IgCONDA-PET (011)’s lead.} The ablation study in which only the spatial-transformer blocks are toggled from $(000) \rightarrow (001) \rightarrow (011)$ shows that self-attention via the spatial transformer is the principal driver of the observed gains by IgCONDA-PET (011). Activating attention at the lowest-resolution already lifts performance (e.g., +3–23 \% on $\lceil \text{DSC} \rceil$ across datasets), and enabling it again at the mid-resolution stage yields a further, comparable increase. Because self-attention mixes information from all spatial tokens, the network learns global anatomical context—comparing a small uptake to homologous tissue elsewhere - and can flag subtle deviations that purely local convolutions miss, all while keeping false positives low.
    \item \textbf{Implicit vs. classifier guidance.} The DPM + CG baseline employs exactly the same diffusion network as IgCONDA-PET (011) but relies on a separate healthy/unhealthy classifier to steer the reverse diffusion. That classifier never sees the highly-noisy intermediates $\bfx_t$, so its gradients are high-variance and can pull the sampler off the PET manifold, re-introducing artifacts or faint lesion remnants. In contrast, IgCONDA-PET’s classifier-free (implicit) guidance re-uses the noise-conditioned diffusion network itself: scaling the difference between its conditional and unconditional scores provides a smooth, internally consistent update at every step, producing cleaner healthy counterfactuals, sharper lesion edges, and fewer false positives.
    \item \textbf{Why plain ViT autoencoding lags behind.} Although VT-ADL shares a ViT encoder with our model, its decoder is a one-shot CNN with neither iterative refinement nor noise-conditioning, and positional information is only approximately restored when tokens are reshaped back to the image grid. Reconstructions are consequently blurred and error maps spill into surrounding tissue. The stochastic, multi-step denoising in IgCONDA-PET peels away healthy anatomy while preserving lesion boundaries, resulting in lower HD95 and higher DSC, AUPRC and detection sensitivity scores.
    \item \textbf{Limitations of f-AnoGAN.} The f-AnoGAN generator performs a single global reconstruction from a latent code; training can collapse to averaged texture and lacks both self-attention and noise-conditioning. Lesion areas often seem to be ``in-painted" with benign texture, reducing residual contrast. IgCONDA-PET avoids this pitfall by iteratively refining a noise-conditioned latent, yielding consistently sharper residuals and better boundary fidelity.
\end{enumerate}

Together, these comparisons pinpoint two synergistic ingredients -- multiscale spatial self-attention via spatial-transformers and implicit diffusion guidance -- as the essential causes of IgCONDA-PET (011)’s consistent superiority over all baselines in our benchmark.

\subsection{Significance testing for performance metrics}
\label{subsec:significance_testing_for_performance_metrics}
\begin{figure}[h]
\centering
\includegraphics[width=0.9\textwidth]{significance_dsc.pdf}
\caption{\textbf{Significance testing for the $\lceil \text{DSC} \rceil$ metric:} Pair-wise Wilcoxon signed-rank tests (Bonferroni-adjusted $\alpha_\text{corr} = 1.78 \times 10^{-3}$) for every pair of methods on each dataset. Blue cells marked $\star$ denote a significant difference ($p < \alpha_\text{corr}$); red cells give the exact $p$-value when the gap is difference between the methods is not significant. IgCONDA-PET (011) is significantly better than all classical baselines on most datasets; excluding the statistical ties with its own ablations (001) and (000), the only statistical ties are with f-AnoGAN on HECKTOR and with VT-ADL on STS, where the nominal DSC difference is small.}
\label{fig:significance_dsc}
\end{figure}

\begin{figure}[h]
\centering
\includegraphics[width=0.9\textwidth]{significance_hd95.pdf}
\caption{\textbf{Significance testing for the HD95 metric:} Pair-wise Wilcoxon signed-rank tests (Bonferroni-adjusted $\alpha_\text{corr} = 2.38 \times 10^{-3}$) for every pair of methods on each dataset. Blue cells marked $\star$ denote a significant difference ($p < \alpha_\text{corr}$); red cells give the exact $p$-value when the gap is difference between the methods is not significant. IgCONDA-PET (011) is significantly better than most classical baselines on most datasets; excluding the statistical ties with its own ablations (001) and (000), the only statistical ties are with f-AnoGAN on HECKTOR, with Classifier+HiResCAM and f-AnoGAN on DLBCL-BCCV, with f-AnoGAN on PMBCL-BCCV, and with f-AnoGAN and VT-ADL on STS, where the nominal DSC difference is small.}
\label{fig:significance_hd95}
\end{figure}

\begin{figure}[h]
\centering
\includegraphics[width=0.9\textwidth]{significance_auprc.pdf}
\caption{\textbf{Significance testing for the AUPRC metric:} Pair-wise Wilcoxon signed-rank tests (Bonferroni-adjusted $\alpha_\text{corr} = 2.38 \times 10^{-3}$) for every pair of methods on each dataset. Blue cells marked $\star$ denote a significant difference ($p < \alpha_\text{corr}$); red cells give the exact $p$-value when the gap is difference between the methods is not significant. IgCONDA-PET (011) is significantly better than most classical baselines on most datasets; excluding the statistical ties with its own ablations (001) and (000), the only statistical ties are with f-AnoGAN on HECKTOR and VT-ADL on STS, where the nominal DSC difference is small.}
\label{fig:significance_auprc}
\end{figure}

\begin{figure}[h]
\centering
\includegraphics[width=0.9\textwidth]{significance_sens.pdf}
\caption{\textbf{Significance testing for the lesion detection sensitivity metric:} Pair-wise Wilcoxon signed-rank tests (Bonferroni-adjusted $\alpha_\text{corr} = 2.38 \times 10^{-3}$) for every pair of methods on each dataset. Blue cells marked $\star$ denote a significant difference ($p < \alpha_\text{corr}$); red cells give the exact $p$-value when the gap is difference between the methods is not significant. IgCONDA-PET (011) is significantly better than most classical baselines on most datasets; excluding the statistical ties with its own ablations (001) and (000), the only statistical ties are with f-AnoGAN on HECKTOR and VT-ADL on STS, where the nominal DSC difference is small.}
\label{fig:significance_sens}
\end{figure}

For every dataset and metric combination, we compared each pair of $n$ methods with a paired Wilcoxon signed-rank test applied to the per-slice (or per-lesion) metric values. The slice-wise pairing controls for inter-patient variability and makes full use of the repeated-measures design. Because, $n=8$ for $\lceil \text{DSC} \rceil$ and detection sensitivity, and $n = 7$ (no Thresholding) for HD95 and AUPRC, there are $n(n-1)/2 = 28$ comparisons for $\lceil \text{DSC} \rceil$ and detection sensitivity, and $n(n-1)/2 = 21$ comparisons for HD95 and AUPRC for each of the datasets. As a result, we applied a Bonferroni correction to base significance level $\alpha = 0.05$ and declared an effect significant when $$p < \alpha_\text{corr} = \frac{0.05}{28} = 1.78\times 10^{-3},$$ for $\lceil \text{DSC} \rceil$ and detection sensitivity, and $$p < \alpha_\text{corr} = \frac{0.05}{21} = 2.38\times 10^{-3},$$ for HD95 and AUPRC. The results of significance testing for the metrics $\lceil \text{DSC} \rceil$, HD95, AUPRC, and detection sensitivity are presented in \Cref{fig:significance_dsc,fig:significance_hd95,fig:significance_auprc,fig:significance_sens}, respectively. The blue cells in \Cref{fig:significance_dsc,fig:significance_hd95,fig:significance_auprc,fig:significance_sens} containing $\star$ represent statistical significance, meaning that the two methods compared are significantly different from one another for that specific metric under the Bonferroni-corrected significance level.

For the $\lceil \text{DSC} \rceil$ metric (see \Cref{tab:dsc_metric}), IgCONDA-PET (011) outperforms all other methods on all datasets except on HECKTOR (where it is beaten by f-AnoGAN by 2.3\%) and on PMBCL-BCCV (where it is beaten by IgCONDA-PET (001) by 0.3\%). But for both these cases, IgCONDA-PET (011) is \textbf{not} significantly different from f-AnoGAN $(p = 0.4483)$ on HECKTOR or IgCONDA-PET (001) $(p = 0.3689)$ on PMBCL-BCCV, as presented in \Cref{fig:significance_dsc}.

Similarly, for the HD95 metric (see \Cref{tab:hd95_metric}), IgCONDA-PET (011) outperforms all other methods on all datasets except DLBCL-BCCV (where it is beaten by IgCONDA-PET (001) by 0.5 pixels). Again, for this case too, IgCONDA-PET (011) is not significantly different from IgCONDA-PET (001) $(p = 0.7259)$.

Furthermore, for the AUPRC metric (see \Cref{tab:auprc_metric}), IgCONDA-PET (011) outperforms all other methods on all datasets except on HECKTOR (where it is beaten by f-AnoGAN by 1.8\%) and on PMBCL-BCCV (where it is beaten by IgCONDA-PET (001) by 1.3\%). Again, for these two cases as well, the differences between IgCONDA-PET (011) and f-AnoGAN on HECKTOR and IgCONDA-PET (011) and (001) on PMBCL-BCCV are not significant with $p=0.0854$ and $p=0.5050$ respectively.

Finally, for the detection sensitivity metric (see \Cref{tab:sensitivity}), while IgCONDA-PET outperforms all other methods on all datasets quantitatively, its performance is not significantly different some of the other methods on various datasets like f-AnoGAN on HECKTOR $(p=0.7494)$, VT-ADL on DLBCL-BCCV $(p=0.0335)$, f-AnoGAN on PMBCL-BCCV $(p=0.0076)$, VT-ADL on STS $(p=0.5328)$, etc. For a complete list of pairwise comparisons that were not statistically significantly different, please refer to the full heatmaps in \Cref{fig:significance_dsc,fig:significance_hd95,fig:significance_auprc,fig:significance_sens}.

Taken together, these results show that IgCONDA-PET (011) delivers the best \textit{average} performance across metrics and datasets, while the isolated quantitative shortfalls are statistically indistinguishable from chance after strict multiple-comparison control. Thus, the occasional statistical tie does not undermine the overall superiority of our method; rather, it highlights the realistic variability one expects across six very heterogeneous PET cohorts.

\subsection{Test set performance as a function of lesion measures}
\label{fig:performance_lesionsize_suvmean_metrics}

\begin{figure}[]
\centering
\includegraphics[width=0.9\textwidth]{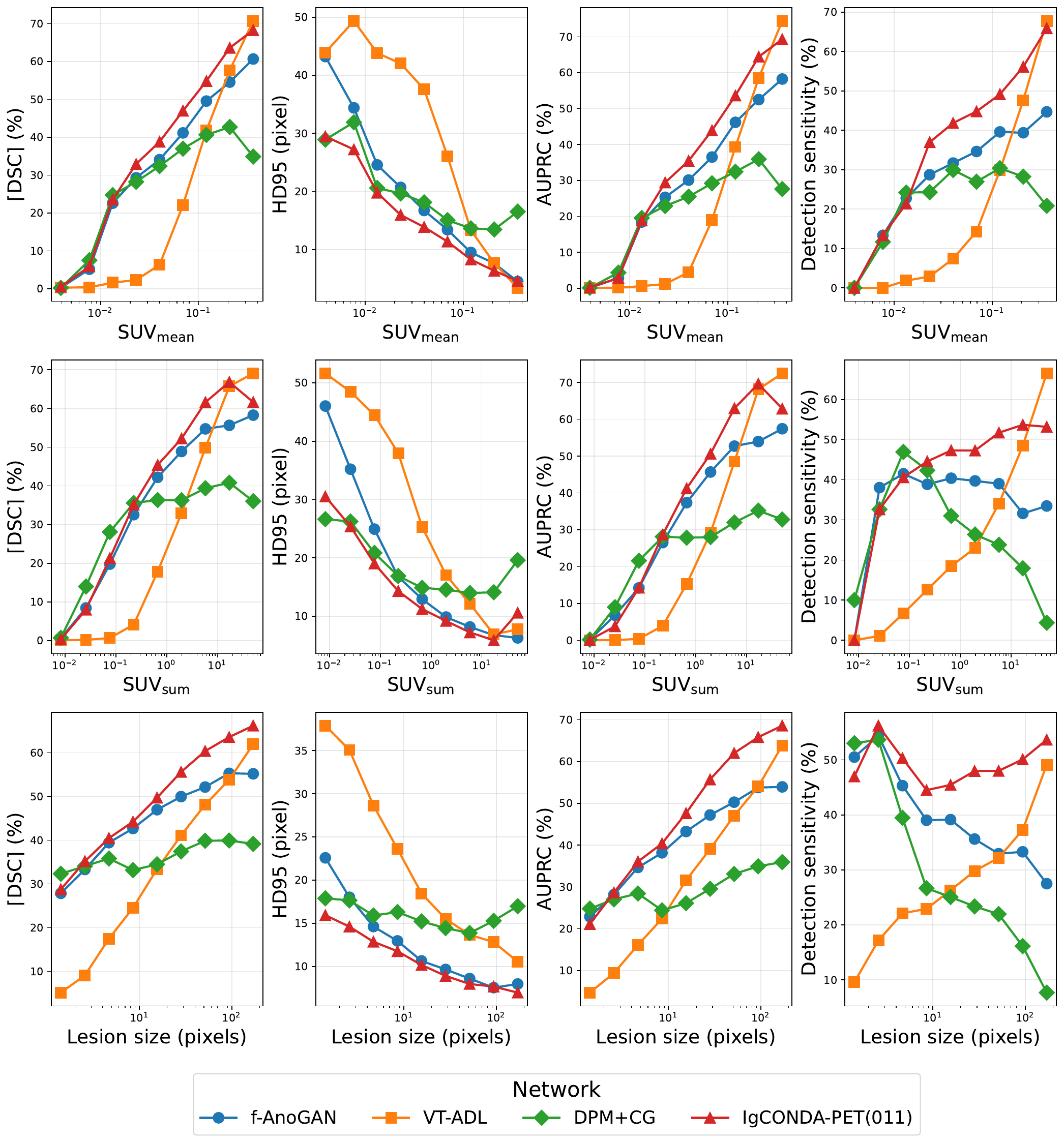}
\caption{Comparison of anomaly detection performance using metrics $\lceil \text{DSC} \rceil$, HD95, AUPRC, and detection sensitivity) stratified by lesion measures, namely SUV$_\text{mean}$, SUV$_\text{sum}$, and lesion size, on the axial slices of PET images. IgCONDA-PET (011) consistently demonstrates superior performance across all metrics and stratification axes compared to baselines (f-AnoGAN, VT-ADL, and DDPM+CG), highlighting its robust generalization capabilities across varying lesion characteristics.}
\label{fig:lesion_measures_vs_metrics}
\end{figure}

\begin{figure}[h]
\centering
\includegraphics[width=\textwidth]{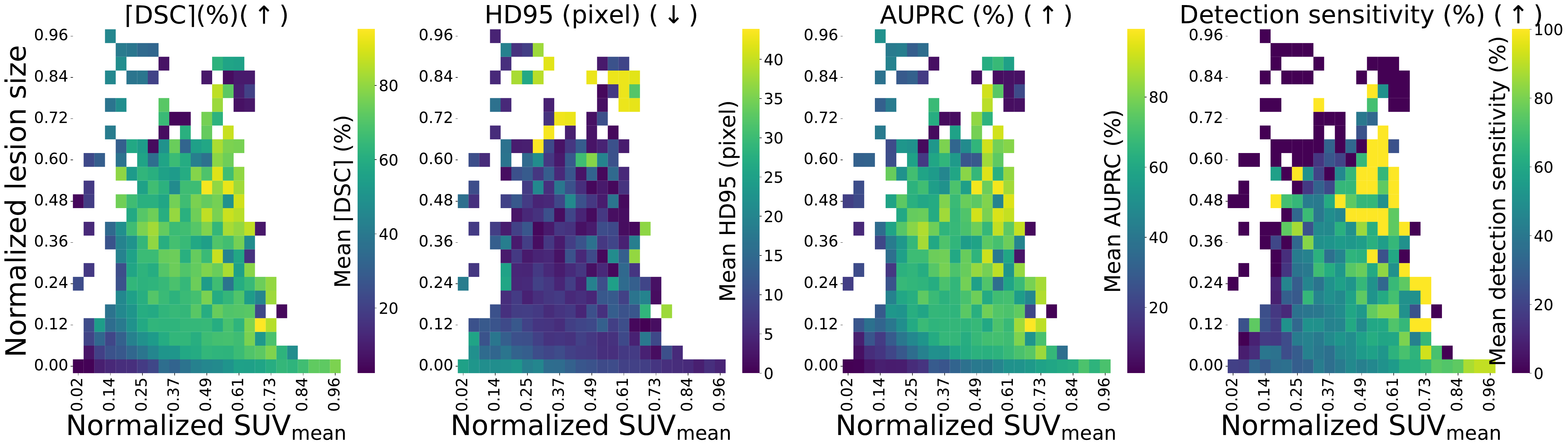}
\caption{Performance on the test set (all internal and external cohorts combined) as a function of normalized SUV$_\text{mean}$ and normalized lesion size on metrics (left to right): $\lceil \text{DSC} \rceil$, HD95, AUPRC, and lesion detection sensitivity. A clear diagonal pattern is visible: metrics improve monotonically from the lower-left corner (tiny, faint lesions) to the upper-right (large, high-uptake lesions), confirming that the diffusion-based model is most reliable on conspicuous foci and still struggles with very small or low-contrast anomalies. For each heatmap, we choose 25 bins for normalized lesion size and SUV$_\text{mean}$. The metric values for all cases falling in each bin were averaged. The empty bins are shown in white.} 
\label{fig:lesionsize_vs_suvmean_vs_metrics}
\end{figure}

In this section, we analyze the performances of various methods as a function of lesion measures, such as SUV$_\text{mean}$ \cite{clinical_metrics_paper, okuyucu2016prognosis}, SUV$_\text{sum}$ \cite{chen2010summation} and lesion size \cite{ahamed2022u} on the axial slices of PET images. These analyses are motivated by two complementary objectives. First, it lets us probe algorithmic robustness against the intrinsic heterogeneity lesion measures within the test set: lesions differ widely in metabolic activity (SUV) and geometric extent, so stratifying results by lesion SUV$_\text{mean}$, SUV$_\text{sum}$ and lesion size reveals whether a method degrades on small, low-uptake foci or only excels on the easy, high-contrast cases. Second, these lesions measures are clinically meaningful prognostic biomarkers in PET-based oncological studies. High SUV$_\text{mean}$ and elevated total lesion glycolysis (approximated here by SUV$_\text{sum}$) correlate with aggressive lesion phenotypes and poorer outcomes, while volumetric burden guides staging and therapy planning. Demonstrating consistent performance across the full spectrum of these biomarkers therefore strengthens the case that a model’s predictions will remain reliable and actionable in real-world clinical decision-making, not just under averaged global metrics \cite{clinical_metrics_paper, dzikunu2025comprehensive}.

As shown in \Cref{fig:lesion_measures_vs_metrics}, across all three lesion-stratification axes, namely SUV$_\text{mean}$, SUV$_\text{sum}$ and lesion size — IgCONDA-PET (011) consistently outperforms every competing baseline (f-AnoGAN, VT-ADL and DDPM+CG) on every evaluation metric considered. It delivers the highest $\lceil \text{DSC} \rceil$, AUPRC, and lesion detection sensitivity and the lowest HD95 in every log-spaced bins (except the bin with the largest mean SUV$_\text{mean}$/SUV$_\text{max}$ where the DSC, AUPRC, and detection sensitivity performances of IgCONDA-PET (011) might degrade slightly as compared to VT-ADL). This uniform dominance indicates that the selective-attention and guidance mechanisms of IgCONDA-PET effectively confer robust generalization across lesion contrasts and scales, resulting in consistently superior segmentation and detection outcomes throughout the entire tumor-burden spectrum in our cohort.

Deep neural networks are often sensitive to lesion size and intensity. Supervised lesion segmentation networks perform better for lesions that are larger and more intense, while failing on very small and faint lesions \cite{clinical_metrics_paper}. We analyze the performance of IgCONDA-PET (011) as a function lesion size and lesion SUV$_\text{mean}$ (both computed in 2D on unhealthy slices). We used a normalized version of lesion size and lesion SUV$_\text{mean}$ for this analyses. For an unhealthy slice, the normalized lesion size was computed as in \Cref{subsec:ablation_over_attention_mechanism_in_different_levels}. Similarly, the pixel intensities of all the slices were normalized in the range [0,1] (as used during training). The mean metric values as a function of normalized lesion SUV$_\text{mean}$ and normalized lesion size are plotted as heatmaps in \Cref{fig:lesionsize_vs_suvmean_vs_metrics} (with 25 bins for both normalized lesion size and normalized SUV$_\text{mean}$). The bins with no slice are colored in white. As seen in \Cref{fig:lesionsize_vs_suvmean_vs_metrics}, the bins along the diagonal (higher normalized lesion size to higher normalized SUV$_\text{mean}$) have a better mean value of metrics within the bins for all metrics. This further confirms that the denoising networks such as diffusion-based IgCONDA-PET (011) adapted for anomaly detection in general also perform better on larger and higher intensity lesions, while unable to detect very small and/or very faint anomalies.  

\subsection{Ablation over attention mechanism in different levels of the network}
\label{subsec:ablation_over_attention_mechanism_in_different_levels}
\begin{figure}[]
\centering
\includegraphics[width=0.95\textwidth]{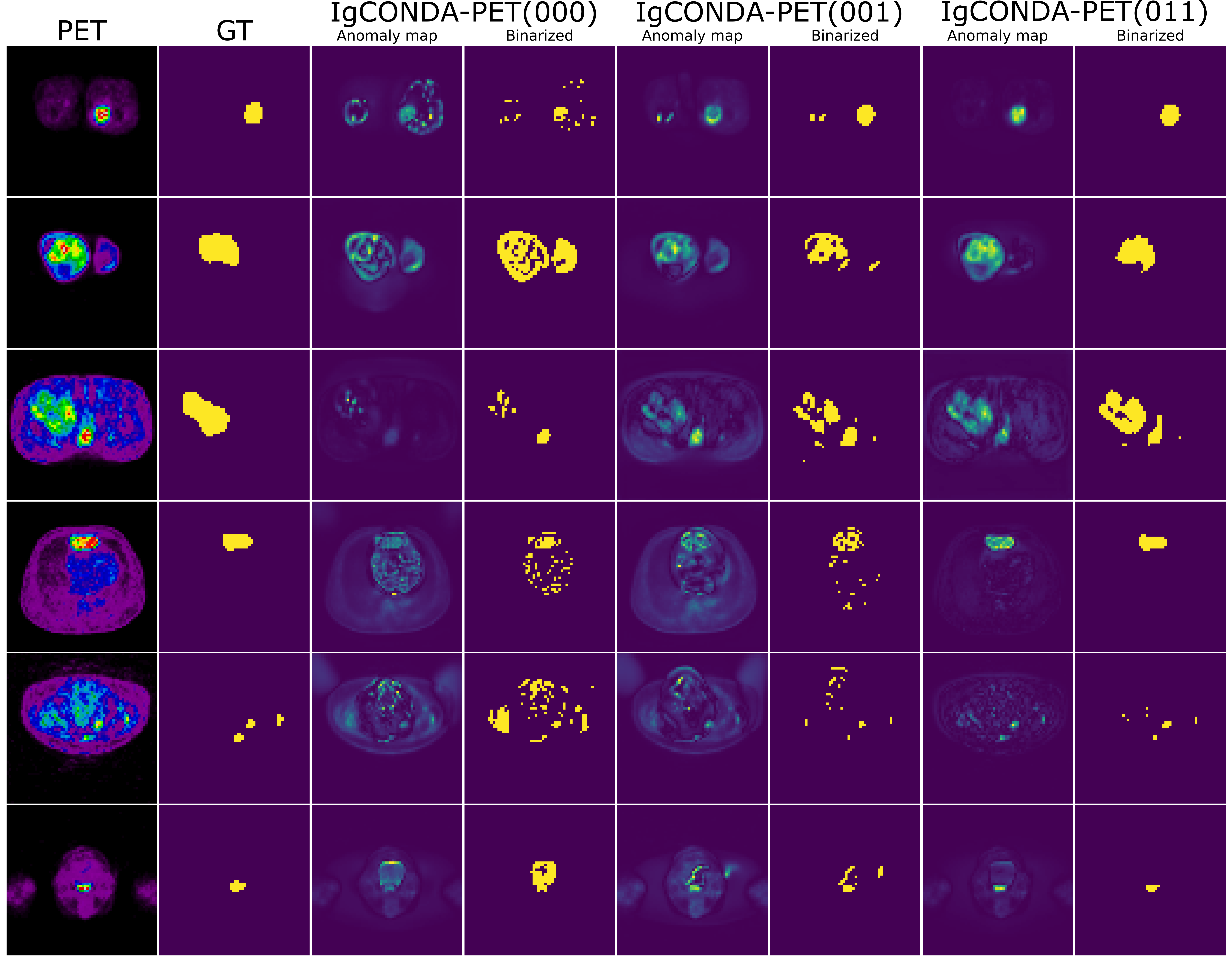}
\caption{Qualitative comparison between the anomaly maps generated by different variants of IgCONDA-PET showing the effect of the presence or absence of attention mechanism in different levels of UNet. For each variant of IgCONDA-PET, we present both the anomaly map (with values in range [0,1]) and its binarized version using the optimal threshold used for computing $\lceil \text{DSC} \rceil$. The IgCONDA-PET (011) variant, which retains spatial transformer in the mid-resolution and lowest-resolution stages of the network, outperforms the other two configurations, producing noticeably sharper and more complete lesion masks with fewer spurious non-pathological activations. In the last two rows, the tiny lesions are clearly delineated only by the (011) variant, highlighting its superior sensitivity to small lesions. Here, the anomaly maps were generated using the inference hyperparameters $D = 400$ and $w = 3.0$ for all variants.}
\label{fig:attention_ablation_viz_labels}
\end{figure}
Attention mechanism typically results in enhanced feature representation which helps the network focus on relevant features in the data by weighting the importance of different areas in the image slices. For PET images, the ability to focus on subtle nuances in pixel intensity and texture variation in the regions of lesions or inflammation is crucial. The attention layers can enhance the network's ability to distinguish these variations from physiological high-uptake regions, thereby improving sensitivity to anomalies. We observed a similar behavior in performance for IgCONDA-PET enhanced with attention mechanism at different levels of the network. From \Cref{tab:dsc_metric,tab:hd95_metric,tab:auprc_metric,tab:sensitivity}, we also notice that, for almost all test sets, IgCONDA-PET (011) outperformed IgCONDA-PET (001), which in turn outperformed IgCONDA-PET (000) (see \Cref{subsec:attention_based_class_conditional_diffusion_model} for a summary on incorporation of class-conditional attention mechanism in the network). In our experiments, we did not ablate over a network with (111)-type attention mechanism (attention in the first level of UNet) since it led to much higher computational costs with almost no improvement in performance. Some representative images showing the anomaly detection performance by the three variants of IgCONDA-PET, (000), (001) and (011) are presented in \Cref{fig:attention_ablation_viz_labels}.

\begin{figure}[h]
\centering
\includegraphics[width=\textwidth]{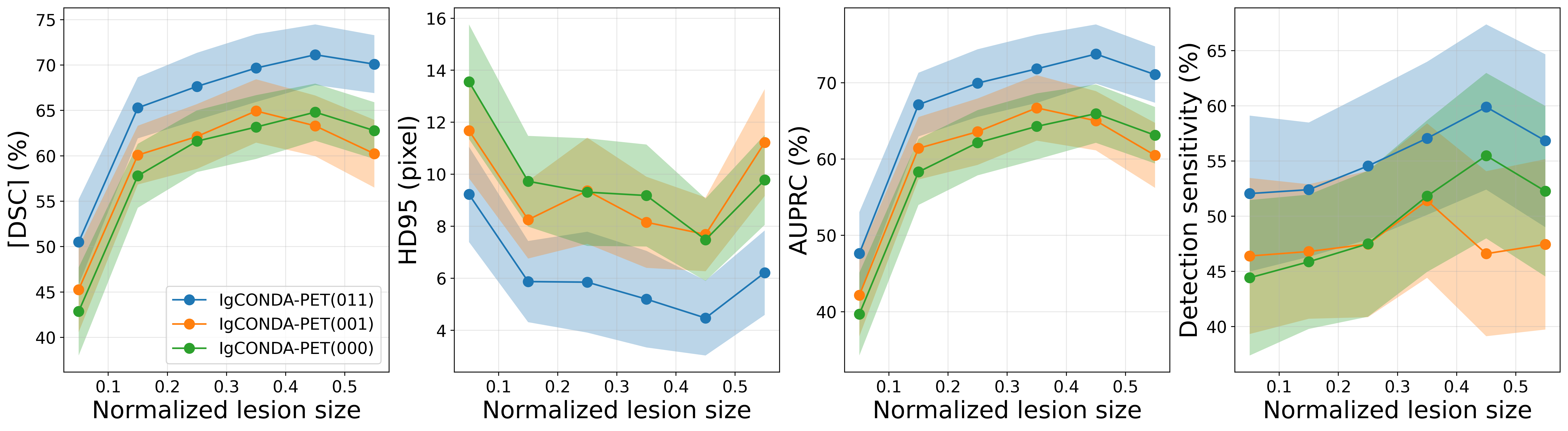}
\caption{Effect of incorporating class-conditional attention via spatial-transformers in the network on anomaly detection performance for small lesions. The small detection performance improves over all metrics, $\lceil \text{DSC} \rceil (\uparrow)$, HD95 $(\downarrow)$, AUPRC $(\uparrow)$, and lesion detection sensitivity $(\uparrow)$, with the incorporation of spatial-transformers in the mid-resolution and lowest-resolution stages of the diffusion network.} Here, the small lesions are defined as the lesions with normalized lesion size < 0.5. The plot shows mean metric in different bins of normalized lesion size along with standard error on mean.
\label{fig:effect_of_attention_on_small_sized_lesions}
\end{figure}

We also analyzed the anomaly detection performance on small lesions for different variants of IgCONDA-PET. To this end, we computed the normalized lesion size for each of the unhealthy slice by counting the number of unhealthy pixels divided by number of unhealthy pixels in the slice with the largest lesion (in 2D) in the test set. Slices with small lesion are defined as those with normalized lesion size values < 0.5. For small lesions slices, the normalized lesion size was binned into 6 bins and the mean and standard error on mean (SEM) values for all metrics were computed in each bin. The results are presented in \Cref{fig:effect_of_attention_on_small_sized_lesions}. We observe that IgCONDA-PET(011) containing attention mechanism within two levels of UNet improved performance for small lesions on metrics $\lceil \text{DSC} \rceil$, HD95 and AUPRC, where the mean $\pm$ SEM for IgCONDA-PET(011) lies well above those for (001) and (000) variants. For the detection sensitivity metric too, the (011) variant had means higher than the other two variants in all bins, although the mean $\pm$ SEM margins were overlapping for all of them.   


\subsection{Sensitivity to inference hyperparameters \texorpdfstring{$D$}{D} and \texorpdfstring{$w$}{w}}
\label{subsec:sensitivity_to_inference_hyperparameteres}

\begin{figure}[h]
\centering
\includegraphics[width=0.5\textwidth]{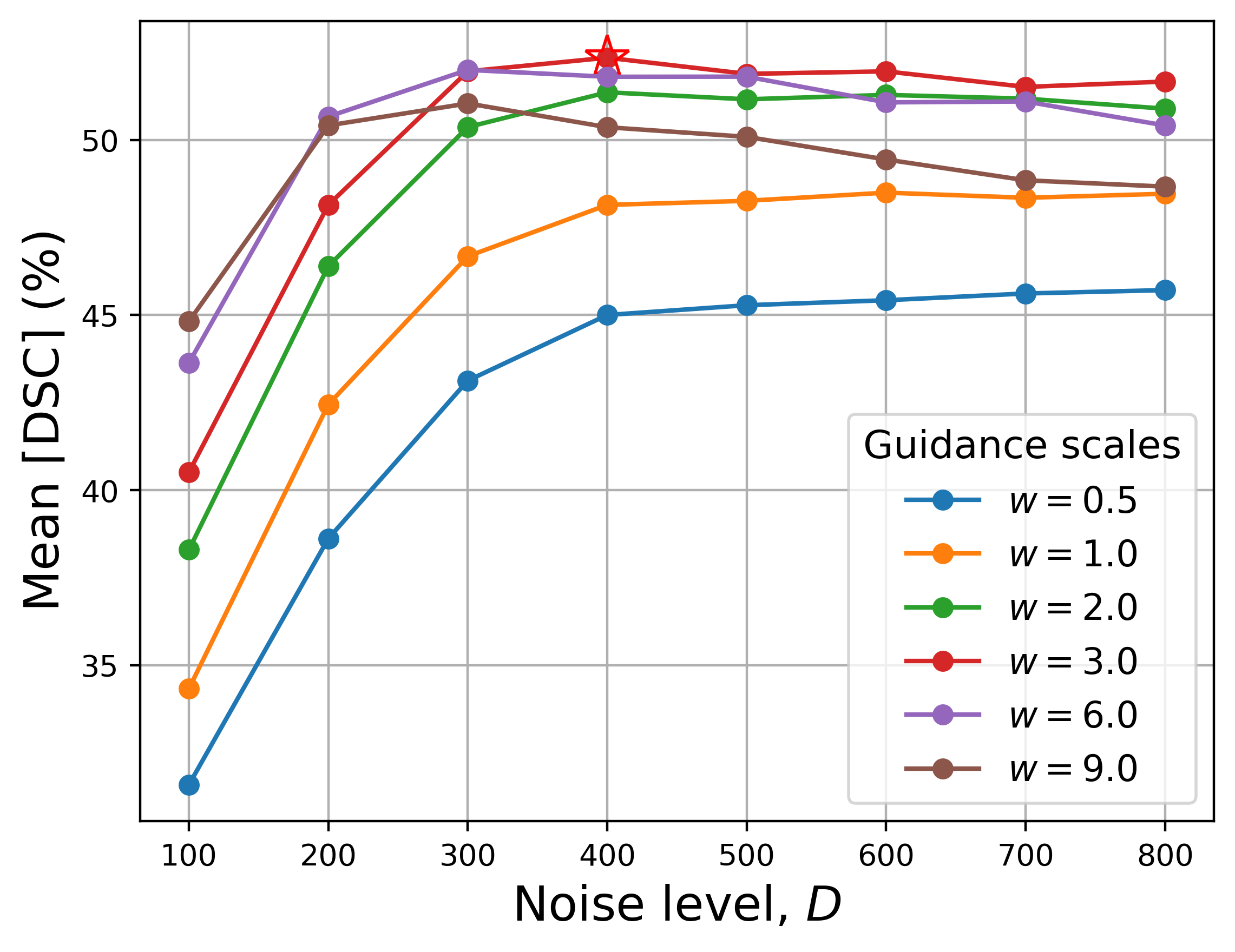}
\caption{Sensitivity to inference hyperparameters, noise level $D$ and guidance scale $w$ for IgCONDA-PET (011). The optimal value for these hyperparameters were chosen on a separate validation set of 10 patients (2 from each of the five internal datasets) consisting of 170 unhealthy slices. The optimal values were found to be $D^* = 400$ and $w^* = 3.0$ which is shown with a red star.} 
\label{fig:inference_hp_sensitivity}
\end{figure}

The optimal values for $D$ and $w$ were obtained after a series of ablation studies using the network IgCONDA-PET (011) on a separate validation data consisting of 10 patients (2 from each of the five internal datasets) consisting of 170 slices. As shown in \Cref{fig:inference_hp_sensitivity}, the mean $\lceil \text{DSC} \rceil$ on the validation set increases with $D$ and then plateaus for all choices of $w$. The optimal mean $\lceil \text{DSC} \rceil$ on the validation set also first increases with increasing $w$ (up to $w=3.0$) and then decreases. From these ablation experiments, we obtained the optimal values as $D^*=400$ and $w^*=3.0$, which were used for performing test set evaluations via implicit guidance on all variants of IgCONDA-PET.

\section{Discussion}
\label{sec:discussion}
In this work, we trained and evaluated IgCONDA-PET, a diffusion model based weakly-supervised anomaly detection framework trained on only image level labels. We employed attention-based class conditioning which were incorporated in different levels of the 3-level denoising UNet and the model was trained on joint conditional and unconditional training objectives. The inference was performed using the reverse of DDIM sampling (with optimal $D^*=400$) by first noise encoding the unhealthy input into a latent representation via the unconditional model and then denoising this latent image using implicit guidance (with optimal $w^*=3.0$) with the conditioning ``healthy'' to generate the corresponding healthy counterfactual (or pseudo-healthy image). The anomaly map was subsequently generated by computing an absolute difference between unhealthy input and healthy counterfactual. Our method with the attention variant (011) outperformed common weakly-supervised/unsupervised anomaly detection baselines by large margins for almost all test sets on metrics such as $\lceil \text{DSC} \rceil$, HD95, AUPRC and lesion detection sensitivity. Using four different types of evaluation metrics, we demonstrated a holistic approach to model performance evaluation at the slice-level ($\lceil \text{DSC} \rceil$ and HD95), pixel-level (AUPRC), and lesion level (lesion detection sensitivity).

It is important to note that since IgCONDA-PET is conditioned to generate healthy counterfactuals from unhealthy images, it effectively does so by reducing the overall intensity of the image in all regions, although the reduction should be much more pronounced in regions of anomalies, as compared to healthy anatomical regions. This helps preserve the healthy anatomical regions, giving rise to more accurate anomaly maps. This makes our model superior to other methods since IgCONDA-PET also has the potential to generate healthy-looking PET images, which are often hard to obtain as patients are usually scanned when there is a possibility for anomalies. Of all the methods explored in this work, only DPM+CG has the potential for counterfactual generation, although it had one of the lowest anomaly detection performances among all the deep learning based methods on our datasets (despite extensive hyperparameter tuning). This method often failed at generating faithful healthy counterfactuals using classifier guidance, giving rise to artifacts in normal anatomical regions leading to higher values in those normal regions on the anomaly maps, as shown in \Cref{fig:compare_different_methods_visualization}. Hence, we show that using classifier-free guidance is superior on our datasets for counterfactual generation as compared to using an extra trained classifier for guidance. \\

We ablated over different attention variants of IgCONDA-PET (000), (001), and (011) and found that incorporating spatial transformers at the last two level of UNet improved performance both quantitatively (see, \Cref{tab:dsc_metric,tab:hd95_metric,tab:auprc_metric,tab:sensitivity}) and qualitatively (see, \Cref{fig:attention_ablation_viz_labels}). Moreover, we also found that the variant (011) improved anomaly detection performance, especially for small lesions (see, \Cref{fig:effect_of_attention_on_small_sized_lesions}). By incorporating attention mechanisms in the last two levels of the network, the model could more effectively integrate and process the higher-level semantic information, which is typically captured in the deeper layers of the network \cite{guo2022attention}. This can be crucial when the distinctions between healthy and unhealthy tissues are subtle, which is often the case in PET images \cite{macpherson2018retrospective}.

Attention mechanisms can also improve the flow of gradients during training, allowing for better and more stable updates. This can result in a more robust learning process, particularly when learning from complex, high-dimensional medical image data \cite{xia2024integrating}. Integrating class embeddings via attention mechanism more deeply into the network likely also allows the model to better use contextual information. This means that the model does not merely look at local features but also considers broader context, which is vital for understanding complex patterns indicative of diseases or other abnormalities. For instance, the presence of a tumor might not only change the texture but also the shape and the relative intensity of the region, which broader contextual awareness can help identify more accurately. 

We also analyzed the anomaly detection performance as a function of normalized lesion size and normalized lesion SUV$_\text{mean}$ and found that IgCONDA-PET performs better on slices containing large and intense lesions than on slices containing smaller and fainter lesions. This is in agreement with several past studies such as \cite{clinical_metrics_paper, xu2023automatic} where the fully-supervised 3D segmentation networks performed better on larger and more intense lesions. 

Small-lesion sensitivity, while already improved by introducing spatial transformers at the mid-resolution and lowest-resolution stages (\Cref{fig:effect_of_attention_on_small_sized_lesions} shows (011) > (001) > (000) across all six size bins) can be pushed further with several complementary strategies, which are all avenues for future work. These include (i) Multi-scale refinement: first running IgCONDA-PET on downsampled $64 \times 64$ images to obtain coarse candidate blobs, then cropping $128–256$ px patches around those blobs and applying a light-weight, high-resolution diffusion refiner, similar to the two-stage scheme proposed in \cite{podell2023sdxl}; (ii) Lesion-aware losses: replace plain MSE with a small-lesion-weighted focal Tversky or Generalized Dice loss so that under-segmenting tiny foci is penalized more heavily than over-segmenting large masses \cite{abraham2019novel,ahamed2023generalized,dzikunu2025adaptive,yeung2022unified}; (iii) Dual-modality guidance: feeding the co-registered low-dose CT as an auxiliary channel so the network can exploit high-frequency anatomical cues for micro-nodules detection \cite{ma2025pet}; and (iv) Small-lesion sensitivity could also benefit from domain-harmonized inputs which incorporate integrating scanner-specific ComBat harmonization \cite{orlhac2022guide} or style-transfer augmentation \cite{zheng2019stada} which are interesting avenues for future work.

Despite outperforming all the baselines, our method (and experimental design) has some limitations. Firstly, it is worth noting that we downsampled axial slices to $64 \times 64$ primarily to (i) fit all baselines into the same GPU memory budget, (ii) keep voxel sizes consistent across images and sites, and (iii) accelerate the iterative diffusion-based models that dominate our training and inference time. The inevitable drawback is a loss of spatial detail: at a typical PET field-of-view ($\sim$40 cm) each pixel covers $\approx$6.3 mm after downsampling, so lesions smaller than two pixels in diameter (<13 mm) are represented by fewer than four voxels. In principle, this can blur low-contrast foci or merge them with background noise, lowering recall for micro-metastases. Additionally, we also performed training on slices of size $128\times128$ (not presented in this paper) which gave a lower anomaly detection performance than on $64\times64$ (and were hence omitted from our analyses). Training diffusion models on high-resolution data can be challenging due to the increased complexity of the image space. Higher resolution images have more details and features, which can complicate the learning process, potentially leading to overfitting or longer training times. Moreover, the problem of anomaly detection from PET is inherently a 3D problem, although training on 3D images would require further downsampling or patch-based approaches \cite{bieder2023memory}. We will explore 3D diffusion-based anomaly detection in our future work. 


Diffusion models for image generation (such as Stable Diffusion) with very high performances are typically pretrained on large datasets consisting of millions of natural images-text pairs \cite{rombach2022high}. Due to the absence of large publicly-available PET datasets of this scale, our diffusion model did not benefit from large-scale pretraining and were all trained from scratch on the datasets used in our work (see, \Cref{subsec:datasets_and_preprocessing}). Recent work \cite{zhang2024diffboost} on using diffusion model for object segmentation in medical images proposed pretraining on RadImageNet dataset \cite{mei2022radimagenet} consisting of 1.35 million radiological images from 131,872 patients consisting of CT, MRI and Ultrasound modalities. Although this dataset does not contain the PET modality, a model pretrained on RadImageNet might still serve as a good starting point for the downstream task of PET anomaly detection, an avenue which can be explored in future work.

The performance of our method is limited by the quality of the generated healthy counterfactuals. Although our method IgCONDA-PET (011) had the best qualitative performance for counterfactual generation among DPM+CG or other attention variants of IgCONDA-PET, there were some cases where the healthy anatomical features were not preserved during the process of conditional decoding to generate healthy counterfactual. For such slices, the generated healthy counterfactuals often failed to align with the image boundaries of the unhealthy inputs, resulting in healthy counterfactuals that appeared significantly different from their expected appearance, thereby diminishing anomaly detection performance. Incorporating additional conditioning signals \cite{zhang2023adding} such as image boundary or edge masks \cite{xie2015holistically} has the potential to further improve the preservation of healthy anatomical features during conditional decoding, thereby also improving the overall anomaly detection performance. Future work will explore this direction, investigating how such enhancements can be systematically integrated into the IgCONDA-PET framework. This approach will not only focus on enhancing the fidelity of the generated images but will also aim to optimize the method’s utility in clinical diagnostic settings, where accurate and reliable anomaly detection is crucial.

\section{Conclusion}
\label{sec:conclusion}

We developed and validated IgCONDA-PET, a diffusion framework for weakly-supervised anomaly detection in PET imaging. Utilizing  attention-based class-conditional diffusion models and implicit guidance, this method efficiently addresses the challenges posed by the scarcity of densely annotated medical images. The counterfactual generation approach, which leverages minimal intervention to translate unhealthy to healthy patient image domains via diffusion noise encoding and conditional decoding, demonstrates remarkable capability in enhancing the sensitivity and precision of PET anomaly detection. Our model not only preserves the anatomical integrity of the generated counterfactuals but also significantly reduces the annotation burden, making it a promising tool for large-scale medical imaging applications. 
\section{Acknowledgment}
\label{sec:acknowledgment}
This work was supported by Canadian Institutes of Health Research (CIHR) Grant PJT-173231. Computational resources were provided by Microsoft AI for Good Lab, Redmond, USA. We acknowledge helpful discussions with Drs.~Yixi Xu and Sara Kurkowska at different stages of this work.

\bibliographystyle{elsarticle-num}
\bibliography{references}

\section*{Statements and declarations}
\renewcommand{\thesubsection}{S\arabic{subsection}}
\subsection{Competing interests}
Nothing to disclose.
\subsection{Ethics approval}
\label{ethical_statements}
The three private data cohorts in this study consisting of PET images of human patients presenting different lymphoma phenotypes. For DLBCL-BCCV and PMBCL-BCCV cohorts, the ethics approval was granted by the UBC BC Cancer Research Ethics Board (REB) (REB Numbers: H19-01866 and H19-01611 respectively) on 30 Oct 2019 and 1 Aug 2019 respectively. For DLBCL-SMHS cohort, the approval was granted by St. Mary's Hospital, Seoul (REB Number: KC11EISI0293) on 2 May 2011 \cite{clinical_metrics_paper}. Due to the retrospective nature of these datasets, the need for patient consent was waived for these three cohorts. These cohorts also comprise of datasets that are privately-owned by the respective hospitals. The public datasets (i) AutoPET 2024 \cite{autopet_paper} was released under CC-BY 4.0 licence which permits unrestricted use, redistribution, adaptation, and commercial exploitation worldwide, provided proper attribution is given and any changes are indicated, (ii) HECKTOR 2022 \cite{hecktor_paper} was distributed under a bespoke End-User Agreement (EUA) which grants a research-use-only, non-commercial, non-redistributable licence, and (iii) STS \cite{vallieres2015radiomics} was shared under CC BY 3.0 which permits copying, redistribution, adaptation, and commercial use of the work, so long as appropriate credit is given to the original creator. All the public datasets were acquired from \textit{The Cancer Imaging Archive}. Their ethical statements can be found in the respective publications. 


\end{document}